\documentclass[aps,pre,twocolumn,showpacs,amsmath,amssymb]{revtex4} %
\usepackage{graphicx}
\usepackage{subfigure}
\usepackage{bm}
\usepackage{color}
\usepackage{amsmath, amssymb,txfonts,siunitx,numprint}

\newcommand{\vc}{\mathbf}

\newcommand{\ave}[1]{\left< #1 \right>}

\newcommand{\del}[3] {\frac{\partial^{#3} #1}{\partial #2^{#3}}}
\newcommand{\dev}[3]{\frac{\mathrm{d}^{#3} #1}{\mathrm{d}#2^{#3}}}
\newcommand{\pdev}[3]{{\mathrm{d}^{#3} #1}/{\mathrm{d}#2^{#3}}}
\newcommand{\intd}[1]{\mathrm{d} {#1}}

\begin{document}
\title{
Negative Strain--Rate Sensitivity in Metallic Glasses Driven by Rejuvenation--Relaxation Competition: Kinetic Monte Carlo Simulations and a Minimal Effective Model
}

\author{ Tomoaki Niiyama$^{1}$ }
\email{niyama@se.kanazawa-u.ac.jp}
\author{Akio Ishii$^{2}$}
\email{ishii@me.es.osaka-u.ac.jp}
\author{
Takahiro Hatano$^{3}$
}
\author{
Tomotsugu Shimokawa$^{1}$
}
\author{
Shigenobu Ogata$^{2}$
}

\affiliation{
${}^1$Faculty of Mechanical Engineering, Kanazawa University, Kakuma-machi, Kanazawa, Ishikawa 920-1192, Japan}
\affiliation{
${}^2$Department of Mechanical Science and Bioengineering, The University of Osaka, Osaka 560-8531, Japan}
\affiliation{
${}^3$Department of Earth and Space Science, The University of Osaka, Osaka 560-8531, Japan}

\date{\today}
\begin{abstract}
When strain-rate sensitivity (SRS) is negative in metallic glasses, the material becomes weaker as the deformation rate increases, leading to accelerated plastic deformation and, eventually, catastrophic fracture.
In this study, we elucidate the mechanism underlying the negative SRS using micromechanics-based kinetic Monte Carlo simulations that couple heterogeneous randomized shear transformation zone (STZ) models for metallic glasses. The model accounted for both the thermomechanical structural rejuvenation and relaxation of the energy barrier for thermal activation of STZs, incorporating a Kohlrausch-Williams-Watts (KWW)-type relaxation function.
The present simulations systematically reproduce the dependence of flow stresses on strain rate, temperature, and the form of the relaxation function.
The SRS tends to decrease at high strain rates and low temperatures in the simulations, and negative SRS appears when a compressed-exponential relaxation function is employed.
Shear localization also appears; however, the conditions under which the observed localization emerges do not fully coincide with those leading to the negative SRS, leaving the dominant factor unclear. 
To clarify the dominant factor, we introduce a simplified theoretical model that reproduces flow stresses consistent with the simulation results. 
An analytical expression derived from the theoretical model reveals that negative SRS originates primarily from the temporal evolution of the activation barrier. Specifically, negative SRS arises when the timescale of external loading exceeds that of STZ relaxation.
\end{abstract}


\maketitle

\section{Introduction}

The dependence of the strain rate on the applied stress
is one of the most significant factors governing the mechanical stability of various materials~\cite{Liang1999CriticalReview,Kang2000CrashSRS,Schuh2007ReviewAmorphousAlloy,Heslot1994CreepStickslipFriction,Scholz1998EarthquakeFriction}.
This stress-strain-rate relationship, called strain-rate sensitivity (SRS), is typically expressed through an index defined as
\begin{align}
 m^\dagger = \partial \ln \bar{\sigma}/\partial \ln \dot{\varepsilon},
\label{eq:m}
\end{align}
where $\bar{\sigma}$ and $\dot{\varepsilon}$ denote 
a representative stress characterizing the flow state and the strain rate, respectively.
The positive SRS is particularly important in solid materials 
because it promotes mechanical stability after yielding (in the flow regime).
On the other hand, the {\it negative} SRS, in which the flow stress decreases as the external loading rate increases, can pose serious challenges for materials.
As the deformation rate increases, the material weakens, 
and the resulting acceleration of the plastic deformation 
can lead to rapid failure of the material.

Metallic glass, a type of amorphous solid, is one of the materials for which this negative SRS is particularly detrimental.
 Despite their high strength and corrosion resistance, metallic glasses often fail catastrophically without warning. This brittleness is partly rooted in their negative SRS; thus, quantifying and controlling this property is critical for the reliable application of metallic glasses.
Another typical example of an amorphous material, apart from metallic glasses,
is fault gouge in seismology.
The SRS, which can be interpreted as the loading rate dependence of friction,
is central to understanding earthquake phenomena
because earthquakes correspond to unstable slip along a fault zone~\cite{Marone1998LabExpFriction,Scholz1998EarthquakeFriction}.

The negative SRS of metallic glasses is mainly observed in deformation tests at low temperatures and high strain rates~\cite{Mukai2002,Li2003nSRS,Lu2003nSRS,DallaTorre2006nSRS,Dubach2007nSRS,Ma2009nSRS,Narasimhan2015nSRS,Zhang2016SRSandAlphaRelaxationMG,Li2016nSRS, Li2017nSRS,Xue2017nSRS,Sahu2019negativeSRSmechanism};
however, the underlying mechanisms and governing factors responsible for this sensitivity remain subjects of ongoing debate, as discussed below.
Recent experimental observations have shown a relationship between negative SRS and the morphology of plastic deformation,
such as vein patterns, nano-corrugations, and shear bands, which has attracted considerable attention~\cite{Mukai2002,Li2003nSRS,Lu2003nSRS,Dubach2007nSRS,Ma2009nSRS,Narasimhan2015nSRS,Zhang2016SRSandAlphaRelaxationMG,Li2016nSRS, Li2017nSRS,Xue2017nSRS,Limbach2017nSRS,Boltynjuk2018nSRS}.
On the other hand, the experimental observation of molten droplets on the fracture surfaces by Li {\it et al.} indicates that  
plastic deformation induced by an increase in local temperature at the shear transformation zone (STZ) gives rise to negative SRS~\cite{Li2017nSRS},
where an STZ is a region consisting of tens to hundreds of atoms that can produce local plastic deformation through atomic rearrangements~\cite{Argon1979STZ}.
The contribution of local heating, as well as that of free volume and potential energy, was also pointed out by a molecular-dynamics simulation study by Yang {\it et al.}~\cite{Yang2018negativeSRSbyMD}.
The mechanism of the negative SRS has been proposed to arise from 
competition between rejuvenation and relaxation of the STZs, based on the similarity between serrated flow behavior and dynamic strain aging~\cite{DallaTorre2006nSRS,Dubach2007nSRS}.
Similarly, Sahu {\it et al.} also demonstrated a relationship between atomic structural relaxation and negative SRS through their nanoindentation tests on metallic glass films combined with molecular dynamics simulations~\cite{Sahu2019negativeSRSmechanism}.
Despite these extensive studies, the dominant factors and the manner in which each contributes to SRS remain unclear,
because exploring a wide range of strain rates and independently controlling individual factors present significant experimental challenges.

A significant theoretical approach to the emergence condition for negative SRS, based on the introduction of a constitutive model, has been proposed by Dubach and coworkers~\cite{Dubach2009modelSRSofBMG}.
They described the condition for negative SRS using a variable $\Delta g$ representing the relaxation of STZs as:
\begin{align}
 \pdev{\Delta g}{\ln \dot{\varepsilon}}{} < - k_B T,
\label{eq:Dubach}
\end{align}
and thus argued that the state of relaxation is the dominant factor governing negative SRS, where $\dot{\varepsilon}$, $k_B$, and $T$ are the strain rate, Boltzmann's constant, and temperature, respectively.
However, verification of this inequality remains elusive because the explicit form of $\Delta g$ and its dependence on the strain rate are undetermined.

Numerical simulation, specifically kinetic Monte Carlo (kMC) simulation, is a promising approach for reproducing negative SRS, while atomic-scale simulations are not well suited for investigating the SRS, because such simulations must span several orders of magnitude in strain rate.
kMC simulations have the potential to overcome this timescale limitation, because the kMC scheme updates the system state only at discrete event times, bypassing the dynamics between successive events.
In the kMC simulation scheme, a metallic glass is coarsely divided into voxels with side lengths of several nanometers, each representing an STZ. 
Each STZ is assigned an activation energy that describes structural rejuvenation (softening) of the STZ by deformation and relaxation (strengthening) over time, and its transformation occurs probabilistically according to transition-state theory~\cite{Zhao2013MGkMCmodel}.
Hence, if appropriate activation energies for STZs in the metallic glasses are specified, SRS can be reproduced through the kMC simulations.

In this study, to theoretically elucidate the primary factor responsible for negative SRS, we derive the activation barrier describing the relaxation behavior of STZs from the time-dependent viscosity of metallic glasses and perform kMC simulations based on micromechanics. These simulations incorporate the spatial heterogeneity of STZ deformation as represented by variations in the energy barrier.
Based on the simulation results, we discuss the condition for the negative SRS by introducing an effective one-body theoretical model that reproduces the kMC results.
In Section~\ref{sec:method}, the details of the simulation scheme are described. We also derive an energy barrier description, which serves as an STZ model for thermally activated deformation and captures both rejuvenation and Kohlrausch-Williams-Watts (KWW)-type relaxation effects.
In Sec.~\ref{sec:results}, the strain-rate dependence of the flow stress obtained from the kMC simulations is presented, including the emergence of negative SRS.
To explain the simulation results, we construct a theoretical model based on a rate description of STZ deformation and derive an explicit expression for the flow stress as a function of strain rate in Sec.~\ref{sec:theo-analysis}.
Finally, a generalized condition for the negative SRS is derived from the expression of the flow stress, and the underlying mechanism of the SRS is discussed in Sec.~\ref{sec:discussion}.

\section{Methodology}
\label{sec:method}

We conducted two-dimensional kMC simulations of metallic glasses based on the method of Zhao {\it et al.}, which captures deformation-induced rejuvenation and time-dependent relaxation~\cite{Zhao2013MGkMCmodel}. A physically motivated activation-energy model and a tailored time-advancement scheme were introduced into their framework.

In these simulations, the metallic glass is divided into $N_x N_y$ voxels with side length $d$ which are treated as STZs, and the position of each voxel is denoted as $\vc{x} = (n_x d, n_y d)$, where $n_x = 1, ..., N_x$ and $n_y = 1, ..., N_y$.
Each voxel, i.e., each STZ, is assumed to transform probabilistically according to a transition rate derived from the activation energy and the local stress $\sigma_{ij}(\vc{x})$ acting on the STZ. 
The activation energy incorporates the effects of rejuvenation (weakening) and structural relaxation (recovery).
The local stress $\sigma_{ij}(\vc{x})$ is the sum of the externally applied load and the stress field determined by micromechanics based on the eigenstrain field~\cite{Mura2013micromechanics,Zhao2013MGkMCmodel}, where $i$ and $j$ are indices representing $x$ or $y$ directions.
The eigenstrain at an STZ is increased by its transformation event, i.e., an elementary process of plastic deformation. In this study, we refer to the increment of the eigenstrain $\Delta \epsilon^{(m)}_{ij}(\vc{x})$ as the transformation strain (or plastic strain), where $m$ denotes the deformation mode [see Fig.~\ref{schem}(a)].

In Sec.~\ref{subsec:ZhaoSTZmodel}, we introduce the fundamental relation between the transition rate of STZ transformation and the time-dependent activation barrier.
The formalism for the time dependence is derived from the viscous behavior of metallic glasses in Sec.~\ref{subsec:KWW_STZ}.
In Secs.~\ref{subsec:kMC} and \ref{subsec:params}, we describe the kMC setup (including our custom procedures) and parameter settings, respectively.

\subsection{Heterogeneously randomized STZ model}
\label{subsec:ZhaoSTZmodel}

We used the heterogeneously randomized STZ model proposed by Zhao and his coworkers~\cite{Zhao2013MGkMCmodel} for the kMC simulations in this study.
The model considers local plastic deformation in the metallic glass as
the transformation (or plastic deformation) of an STZ, and
assumes that the transformation is governed by the thermal activation of local (or heterogeneous) randomized STZ plastic deformation events.

\begin{figure}[tbp]
 \begin{center}
\includegraphics[bb=6 8 441 460,width=\linewidth]{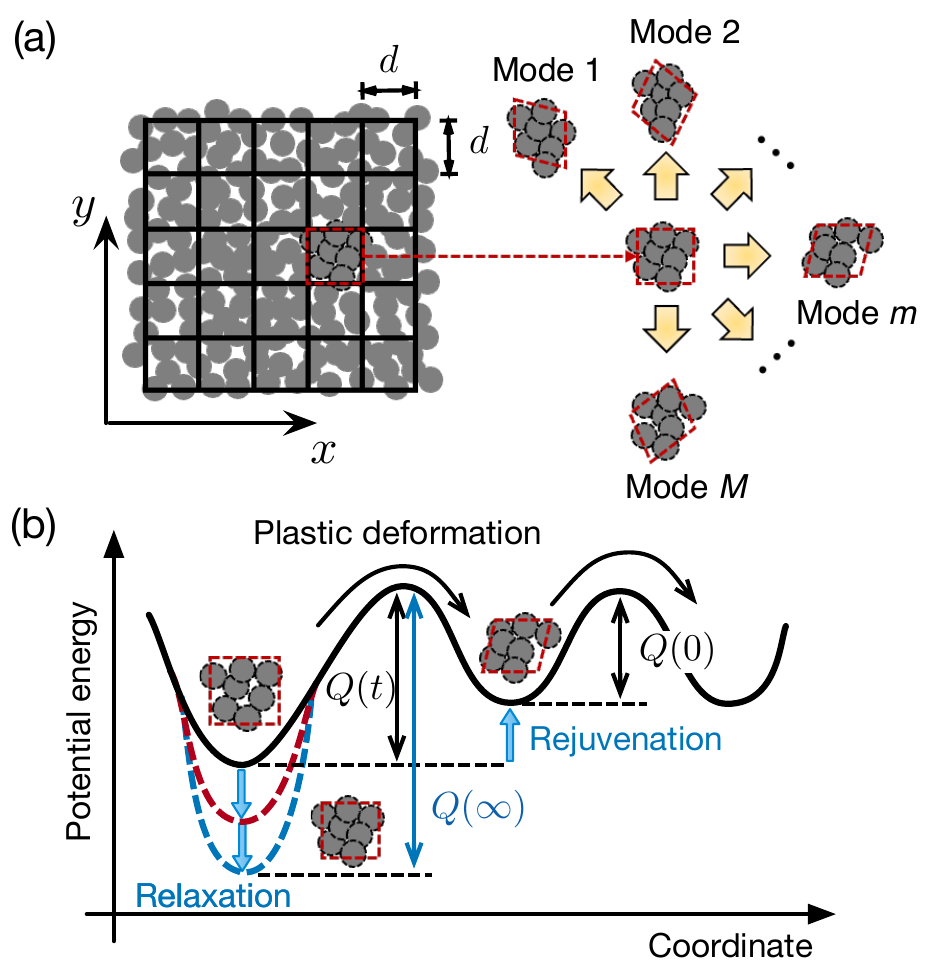}
 \caption{Schematic of the heterogeneously randomized STZ model. (a) Each STZ possesses multiple plastic deformation modes with randomized shear strains.
(b) Potential energy surface and changes in the energy barrier during STZ deformation (plastic deformation) and structural relaxation. }
 \label{schem}
 \end{center}
\end{figure}

As shown in Figure~\ref{schem}(a), we consider the metallic glass 
as a two-dimensional system represented 
by the coordinate $\vc{x} = (x, y) = (n_x d, n_y d)$.
The material is treated as a continuum body filled with voxels representing STZs, each of which possesses $M$ multiple plastic deformation modes with randomized transformation shear strains.
According to the transition-state theory,
the $m$-th plastic deformation mode of an STZ at the position 
$\vc{x}$ is thermally activated with the transition rate:
\begin{align}
 k^{(m)}(\vc{x}) = k_0 \exp \left[ - G^{(m)}(\vc{x})/k_B T \right],
\label{eq:rate}
\end{align}
where $k_0$, $k_B$, and $T$ are the frequency factor, Boltzmann's constant,
and temperature, respectively.
The effective (stress-biased) activation barrier $G^{(m)}(\vc{x})$ can be represented by the following equation~\cite{Argon1996DeformCrystalBook,Zhao2013MGkMCmodel}:
\begin{align}
G^{(m)}(\vc{x}) &= Q^{(m)}[\vc{x}, t(\vc{x})]
- \frac{1}{2} \sum_{i, j} 
\sigma_{ij}(\vc{x}) \Delta \epsilon^{(m)}_{ij}(\vc{x}) V_\mathrm{STZ},
 \label{eq:act_E}
\end{align}
where the tensor $\Delta \epsilon_{ij}^{(m)}$ represents 
the transformation strain produced by the deformation with mode $m$, 
$\sigma_{ij}(\vc{x})$ is the local stress tensor at position $\vc{x}$,
and $V_\mathrm{STZ}$ is the volume of the STZ corresponding to a voxel.
We denote by $Q^{(m)}(\vc{x}, t)$ the intrinsic (stress-free) energy barrier between an initial and a saddle-point configuration, which corresponds to the time-dependent energy barrier for thermally activated STZ deformation in mode $m$ as illustrated in Fig.~\ref{schem}(b). 
The variable $t(\vc{x})$ (or $t$) is the elapsed time since the previous plastic deformation of the STZ at $\vc{x}$.
This elapsed time is utilized to describe the rejuvenation 
and relaxation processes of STZs. 

The effects of the rejuvenation (softening) by deformation and 
relaxation (strengthening) over time can be interpreted in terms of 
the behavior of the intrinsic energy barrier.
In Fig.~\ref{schem}(b), we show a schematic of the potential energy profile during the plastic deformation of a representative STZ.
After an STZ undergoes deformation, it falls into a basin of the potential energy surface~\cite{Liu2018PESofMG}.
The energy of this state (the energy of the local minimum within the basin) 
generally increases, corresponding to rejuvenation.
This deformation-induced rejuvenation lowers the barrier height
for the subsequent STZ deformation, denoted as $Q(0)$, thereby facilitating further deformation (softening).
As time $t$ elapses, the STZ structure relaxes (or ages), lowering the energy of the metastable states and increasing the energy barrier for STZ plastic deformation, denoted as $Q(t)$~\cite{Zhao2013MGkMCmodel,Qiao2016StressRelaxationMG}. 
This relaxation process 
($\beta$ relaxation~\cite{Luo2017RelaxationMG}) makes the STZ progressively harder to transform over time (strengthening).

To implement the processes described above, the relaxation process is modeled through the time-dependent energy barrier $Q^{(m)}[\vc{x}, t(\vc{x})]$, while the rejuvenation process is modeled by resetting the elapsed time $t$ to zero whenever the STZ at $\vc{x}$ undergoes plastic deformation.

\subsection{Time-dependent energy barrier: description of structural relaxation}
\label{subsec:KWW_STZ}
For the present simulations, we employ a time-dependent intrinsic energy barrier consistent with a well-known characteristic of metallic glasses: their viscosity evolves over time due to structural relaxation, instead of using the energy barrier adopted in the work of Zhao {\it et al}.
In this subsection, we derive the time-dependent energy barrier, which we denote as $Q(t)$ for convenience, from the time-dependent viscosity $\eta(t)$ of metallic glasses.

It is well known that the viscosity of metallic glasses changes as time elapses due to structural relaxation and can be described by a KWW function~\cite{Busch1998KWW}:
\begin{eqnarray}
\eta(t) = \eta_0 + 
\delta \eta \left( 
  1 - \exp \left[ -\left( \frac{t}{t_R} \right)^{\beta} \right] 
\right),
 \label{eqn:relax}
\end{eqnarray}
where $\eta_0$ is the viscosity before relaxation. 
The exponent $\beta$ is a positive parameter that characterizes the dynamic heterogeneity of the relaxation and the fundamental behavior of the process~\cite{Richert2002heteroDynamicsLiquid}.
Although $\beta <1$ has traditionally been assumed for metallic glasses, corresponding to {\it stretched-exponential relaxation}~\cite{Phillips1996StretchedExpRelax,Ishii2021KWWStressRelax},
some recent studies have reported {\it compressed-exponential relaxation} ($\beta > 1$) associated with gel-like behavior~\cite{Morishita2012CompressedExpRelax,Ruta2012CompressedExpRelaxMG,Ruta2013CompressedExpRelaxMG,Luo2017RelaxationMG}.

The parameter $\delta\eta$ represents the change in viscosity
and can be approximated by the fully relaxed viscosity $\eta(\infty)$,
which is several orders of magnitude larger than $\eta_0$~\cite{Busch1998KWW}. 
The time constant $t_R$ is the characteristic relaxation time, representing the average time for relaxation, typically expressed in Arrhenius form: 
\begin{eqnarray}
t_R
= \frac{1}{k_0} \exp \left( \frac{Q_R}{k_{\rm B}T} \right),
\label{eq:t_R}
\end{eqnarray}
where $Q_R$ is the activation energy of the relaxation process.

The viscosity $\eta$ can also be related to 
the time-dependent energy barrier of STZ deformation $Q(t)$ 
via Andrade's equation~\cite{Bird2002TransportPhenomena}:
\begin{eqnarray}
\eta(t) = {\rm A}\exp\left(\frac{Q\left(t \right)}{k_{\rm B}T}\right),
 \label{eqn:eta}
\end{eqnarray}
where $A$ is a constant.
To connect the equation to Eq.~(\ref{eqn:relax}), we consider that $Q(t)$ consists of a time-independent term $Q_0$ and a time-dependent term $\Delta Q(t)$;
\begin{eqnarray}
Q(t)\equiv Q_0+\Delta Q(t).
\label{eq:Q_t} 
\end{eqnarray}
Assuming that $\eta_0$ corresponds to the term including $Q_0$ as follows:
\begin{eqnarray}
\eta_0 \approx {\rm A}\exp\left(\frac{Q_0}{k_{\rm B}T}\right), 
\end{eqnarray}
we can express the viscosity in terms of the time-dependent activation energy as:
\begin{eqnarray}
\eta(t) = \eta_0 \exp \left[ \frac{\Delta Q(t)}{k_{\rm B}T} \right].
\end{eqnarray}
Combining this equation with Eq.~(\ref{eqn:relax}),
we obtain the time-dependent activation barrier $\Delta Q(t)$ as
\begin{eqnarray}
\Delta Q(t) = 
k_{\rm B}T \ln \left[ 1+ \frac{\delta\eta}{\eta_0} 
\left\{ 1 - e^{ -\left({t}/{t_R} \right)^{\beta}} \right\} \right]. 
 \label{eqn:deltq}
\end{eqnarray} 
By fitting the relation between temperature and viscosity before and after equilibration, the parameter $\delta \eta / \eta_0$ was determined as
$\delta \eta / \eta_0 = \exp(E_v / (k_{\rm B} T)-60.45)$, where $E_v = 3.28$ eV~\cite{Busch1998KWW}.
In the relevant temperature range
from $T = \SI{300}{K}$ to $\SI{500}{K}$, the parameter is confined
to $10^{7} \lesssim \delta \eta / \eta_0 \lesssim 10^{29}$.
The activation energy for the relaxation time was reported as
$Q_R = 1.2$ eV~\cite{Yu2012betaRelaxMG}.

\begin{figure}[tbp]
 \includegraphics[bb=10 12 278 206,width=6cm]{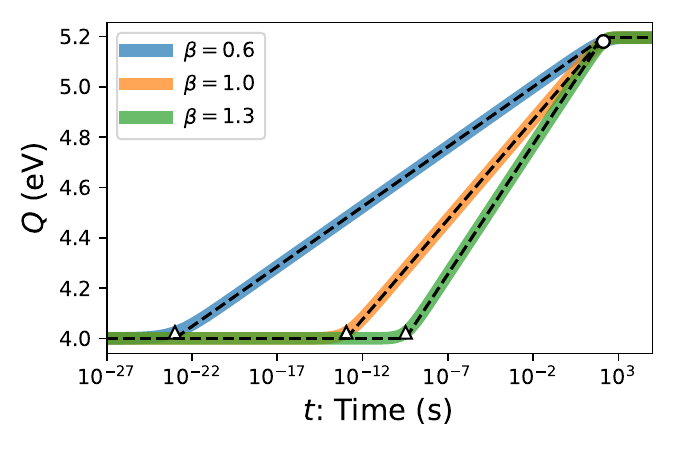}
\caption{
Typical examples of the activation barrier $Q(t)$ at $T=\SI{400}{K}$,
where blue, orange, and green curves correspond to the cases with $\beta = 0.6, 1.0$, and $1.3$, respectively.
The dashed lines describe the approximated shape of $Q(t)$ [Eqs. (\ref{eq:Q_approx-before}), (\ref{eq:Q_approx-trans}), and (\ref{eq:Q_approx-after})].
The circles and triangles represent the barrier heights at $t_R$ and $t_s$, respectively.
}
\label{fig:Q_t}
\end{figure}
Three representative curves of $Q(t)$ at $T=400$ K with different $\beta$ 
are depicted in Fig.~\ref{fig:Q_t}.
All three curves begin to rise around a characteristic time $t_s$ (triangles) and saturate around $t_R$ (circle), where the details of $t_s$ are explained below.
The exponent $\beta$ determines the characteristic time $t_s$ and the rate at which $Q(t)$ increases.
Note that since $Q(t)$ increases with time, the metallic glass described here strengthens as time elapses unless a deformation event occurs.

The specific time $t_s$ is obtained from an approximate expression for $Q(t)$.
Fig.~\ref{fig:Q_t} shows that the temporal evolution of $Q(t)$ can be divided into three regimes: pre-relaxation, transition, and post-relaxation.
In the pre-relaxation regime ($t < t_s$),
the barrier height can be considered constant:
\begin{align}
 Q(t) \simeq Q_0 \quad (t < t_s).
\label{eq:Q_approx-before}
\end{align}
In the transition regime ($t_s < t < t_R$), $Q(t)$ increases with time.
In this regime, considering two approximate expressions,
$ 
\ln \delta \eta / \eta_0 + \ln  (\eta_0/\delta \eta + 1 - e^{-(t/t_R)^\beta})
\simeq \ln \delta \eta / \eta_0 + \ln  (1 - e^{-(t/t_R)^\beta})
$
obtained by applying $\eta_0 / \delta \eta \ll 1$ to Eq.~(\ref{eqn:deltq}),
and 
$1 - e^{-(t/t_R)^\beta} \simeq (t/t_R)^\beta$,
we can describe the barrier height variation as 
\begin{align}
Q(t) \simeq  Q_0 + 
k_B T \left( \beta \ln t/t_R + \ln  \delta \eta / \eta_0 \right)
\quad (t_s < t < t_R).
\label{eq:Q_approx-trans}
\end{align}
The variation after the transition regime is represented by
\begin{align}
 Q(t) \simeq Q_0 + k_B T \ln (1 + \delta \eta / \eta_0) \quad (t > t_R).
\label{eq:Q_approx-after}
\end{align}
The specific time is evaluated as the time satisfying the relation:
$k_B T \left( \beta \ln t_s/t_R + \ln  \delta \eta / \eta_0 \right) = 0$.
Thus, we obtain
\begin{align}
 t_s =  \left( \delta \eta / \eta_0 \right)^{-1/\beta} t_R.
\label{eq:t_start}
\end{align}
The good agreement between the dashed lines shown in Fig.~\ref{fig:Q_t}, representing the approximate expressions of $Q(t)$, and the actual $Q(t)$ indicates the validity of the approximation.
In the figure, the triangle and circle symbols represent $Q(t_s)$ and $Q(t_R)$, respectively.
The above approximations will be used in Sec.~\ref{sec:discussion}.

\subsection{kMC simulation scheme}
\label{subsec:kMC}

We conducted kMC simulations under constant strain-rate conditions using the activation barrier derived in the previous subsection [Eqs.~(\ref{eq:Q_t}) and (\ref{eqn:deltq})].
As the initial condition of the simulations, small random initial eigenstrains, $\epsilon_{ij}(\vc{x})$, were assigned to each STZ site, while preserving the site volumes:  $\epsilon_{xx}(\vc{x}) = - \epsilon_{yy}(\vc{x})$.
These initial eigenstrains represent a random internal stress field arising from the disordered atomic structure peculiar to amorphous solids.
The elapsed times at all sites, $t(\vc{x})$, were set to infinity to ensure that the intrinsic energy barriers are fully relaxed: $Q[\vc{x}, t(\vc{x})] = Q_0 + k_B T \ln (1 + \delta \eta/\eta_0)$.
At this stage, if any site $\vc{x}'$ with mode $m'$ has a negative activation barrier height ($G^{(m')}(\vc{x}') \le 0$), all such sites are transformed according to the procedure described below until all the heights become positive.
Although these initial transformations make the elapsed times at the sites zero, the elapsed times were then reset to infinity to define the initial state.

After the initial preparation, an external loading stress was applied along the $x$-direction to achieve the deformation condition with constant strain rate $\dot{\varepsilon}$.
The local stress on an STZ at $\vc{x}$, $\sigma_{ij}(\vc{x})$, is given by the sum of the internal stress $\sigma^*_{ij}(\vc{x})$ and the externally applied stress. The internal stress field $\sigma^*_{ij}(\vc{x})$ satisfies static mechanical equilibrium and can be calculated by micromechanics based on the eigenstrain field $\epsilon_{ij}(\vc{x})$, namely, the accumulated transformation strain~\cite{Mura2013micromechanics, Zhao2013MGkMCmodel}.
Hence, the local stress along the $x$-direction is expressed as:
\begin{align}
 \sigma_{xx}(\vc{x}) 
= \sigma^*_{xx}(\vc{x}) +  E \dot{\varepsilon} t',
\end{align}
where $E$ and $t'$ are the Young's modulus and the global simulation time, respectively.
Note that $t'$ represents the time elapsed from the beginning 
of the simulations and should be clearly distinguished from $t(\vc{x})$, the elapsed time since the last event at $\vc{x}$.
Under the stress field, $\sigma_{xx}(\vc{x})$, an STZ transforms probabilistically, with its position $\vc{x}'$ and deformation mode $m'$ determined by the procedure explained below.
We refer to this transformation, together with the subsequent changes it induces, as a ``deformation event'' in this study.
Each transformation increases the eigenstrain at the corresponding position by $\Delta \epsilon^{(m')}_{ij}(\vc{x}')$:
\begin{align}
 \epsilon_{ij}(\vc{x}') \to \epsilon_{ij}(\vc{x}') 
+ \Delta \epsilon^{(m')}_{ij}(\vc{x}').
\label{eq:update_eigen_strain}
\end{align}
This modification of the eigenstrain field generates a new internal stress field, $\sigma^*(\vc{x})$, which is recalculated using micromechanics.
After updating the stress field, the next deformation event is determined.
Repeating this procedure forms the main loop of the present kMC simulation.

While our simulation scheme is based on the method of Zhao {\it et al.}, we implemented several modifications in the update scheme, including a backtracking time-advance scheme, continuous eigenstrain increments, and avalanche evolution, as described below.
To generate deformation events probabilistically, we employed the rejection-free algorithm~\cite{Bortz1975rejectionFreeKMC,Gordon1968Adsorption-Isotherms,Abraham1970VaporDeposition,Kohli1972CrystalDissolution,Voter2007kMC_review}  as in the work by Zhao {\it et al.}
The algorithm provides the inter-event interval $T_\mathrm{RF}$ between successive events as:
\begin{align}
 T_\mathrm{RF} &= - (\ln \xi) / k_\mathrm{tot}
\\
k_\mathrm{tot} 
&= \sum_{n_x=1}^{N_x} \sum_{n_y=1}^{N_y} \sum_{m=1}^M k^{(m)}(\vc{x}, t),
\end{align}
where $\xi$ is a random number sampled from a uniform distribution: $0 \le \xi < 1$.
The site $\vc{x}'$ and the mode $m'$ where the transformation occurs are determined according to the probability $k^{(m')}(\vc{x}', t)/k_\mathrm{tot}$.

The standard rejection-free algorithm assumes time-independent transition rates and positive activation barriers.
However, under the constant-strain-rate condition used in this study, these assumptions can be violated: the external load increases linearly in time, so the resulting transition rates are nonstationary; moreover, if a long $T_\mathrm{RF}$ is drawn, the load may increase enough that some intrinsic activation barriers, $G^{(m)}[\vc{x}, t'(\vc{x})]$, become nonpositive before the scheduled event.
To avoid the former issue, we rejected the events if $T_\mathrm{RF} > T^*_\mathrm{RF}$ in our kMC simulations, where $T^*_\mathrm{RF}$ is a parameter taken short enough that the external stress, $E \dot{\varepsilon} t'$, can be considered constant.
For the latter issue, we introduced a backtracking time-advance scheme: if any sites have nonpositive activation barriers at the candidate event time $t' + T_\mathrm{RF}$,
we reject the candidate event and reduce the proposed time advance by half. 
That is, we advance the current global time by $T_\mathrm{RF}/2$ without excuting any event:
\begin{align}
  t' \to  t' + T_\mathrm{RF}/2.
\end{align}
We then redraw a new candidate event (and waiting time $T_\mathrm{RF}$) using the rejection-free algorithm.
Further, if any sites with nonpositive barrier heights remain after the new draw, we repeat this backtracking procedure recursively until the barrier heights at all sites become positive: $G^{(m)}[\vc{x}, t'(\vc{x})] > 0$.
This scheme ensures that all events result from thermal activation processes.

When a deformation event at site $\vc{x}'$ with mode $m'$ is accepted, the eigenstrain of the site increases according to Eq.~(\ref{eq:update_eigen_strain}); however, this increment introduces spatial discontinuities in the eigenstrain and resultant stress field.
To mitigate these discontinuities, we modified the incremental rule of the eigenstrain: the incremental eigenstrain, $\Delta \epsilon_{ij}(\vc{x}')$, generated by the deformation event is distributed to a $3 \times 3$ neighborhood centered on $\vc{x}'$ with the weights assigned as follows:
\begin{align}
 \epsilon_{ij}(\vc{x}') &\to
 \epsilon_{ij}(\vc{x}') + \Delta \epsilon_{ij}(\vc{x}') / 4,
\\
 \epsilon_{ij}(\vc{x}' + \vc{d}_1) &\to 
\epsilon_{ij}(\vc{x}' + \vc{d}_1) + \Delta \epsilon_{ij}(\vc{x}') / 8,
\\
 \epsilon_{ij}(\vc{x}' + \vc{d}_2) &\to 
\epsilon_{ij}(\vc{x}' + \vc{d}_2) + \Delta \epsilon_{ij}(\vc{x}') / 16,
\end{align}
where $\vc{d}_1 \in \{(\pm d, 0), (0, \pm d)\}$ and 
$\vc{d}_2 \in \{ (\pm d, \pm d)\}$ (independent signs).

Further, we introduced a protocol into the simulation to represent the avalanche behavior of the STZ transformation.
After the eigenstrain increment, the stress field $\sigma^*(\vc{x})$ is updated. The update sometimes leads to another site $\vc{x}''$ with mode $m''$ having a negative barrier height: $G^{(m'')}(\vc{x}'') < 0$.
In our protocol, such a site $\vc{x}''$ is treated as a site whose transformation is triggered by the event at $\vc{x}'$, i.e., a chain reaction in an avalanche.
All such sites with a negative barrier height are transformed, and then the stress field is updated. If other negative sites appear again even after the update, the same treatment is conducted recursively until the barrier heights at all sites become positive.
At each update of the protocol, we advance the simulation clock by a short interval $T_\mathrm{defo}$ on the order of an atomic vibrational period. This interval is a characteristic time required for an STZ to deform, which we estimated using the elastic wave transit time across the STZ.
When the barrier heights of all sites are positive, the deformation event is completed, and the next deformation event is drawn by our modified rejection-free algorithm.
Note that this chain reaction of STZ transformation corresponds to the avalanche dynamics in crystalline and amorphous plasticity~\cite{Miguel2001Intermittent-di,Maloney2004AmorphousAvalanches}.

\subsection{Parameters of the simulations}
\label{subsec:params}

For the STZ model and the structural relaxation parameters, we utilized experimental values, including the viscosity of Zr-based metallic glasses.
For Young's modulus and Poisson's ratio, $E = \SI{88.6}{GPa}$
and  $\nu = 0.371$ were employed~\cite{Bian2002MGCNT}.
Note that the deviation in the elastic constants (Young's modulus and Poisson's ratio) of Zr-based metallic glasses with respect to the change in its composition is approximately 10 $\%$~\cite{Wang2012MGProperty}.

The domain of the metallic glass was divided into $256 \times 256$ voxels
with side length $d = \SI{1.7}{nm}$, and the periodic boundary condition was applied.
The total number of deformation modes of each STZ was set to $M = 20$.
The time required for a transformation of an STZ, $T_\mathrm{defo}$, was estimated as the propagation timescale of the elastic wave across three voxels:
$T_\mathrm{defo} = (3d/2)/v = \SI{0.52}{ps}$,
where the velocity of the elastic wave is $v = \SI{5e+3}{m/s}$.
The initial eigenstrain of each voxel, $\epsilon_{ij}(\vc{x})$, was assigned from a normal distribution with mean zero and standard deviation $0.005$.

The intrinsic energy barrier of an STZ transformation in mode $m$ at the position $\vc{x}$ is described by 
$Q^{(m)}(\vc{x}) \equiv Q^{(m)}_0(\vc{x}) + \Delta Q[t(\vc{x})]$, 
where $\Delta Q[t(\vc{x})]$ is the KWW-type relaxation model 
described by Eq.~(\ref{eqn:deltq}), 
and $t(\vc{x})$ is the elapsed time since the previous deformation event 
at the site $\vc{x}$.
Note that $\Delta Q(t)$ is independent  of the deformation mode
because the relaxation or rejuvenation feature is related to the individual atomic structure of an STZ rather than its deformation mode.
To represent the energetic heterogeneity of STZs,
which is due to the differences in the atomic structure 
and local stress~\cite{Srolovitz1981RDFStrctRelaxGlass,Launey2008MG,Egami2011AtomicStress},
the time-independent term, $Q^{(m)}_0(\vc{x})$, of all deformation modes 
in each STZ was sampled from a normal distribution with a mean of $4.0$ eV 
and a standard deviation of $0.8$ eV.  
For the characteristic exponent $\beta$ in $\Delta Q[t(\vc{x})]$,
we employed three typical values:
$\beta=0.6$ (stretched exponential relaxation)~\cite{Phillips1996StretchedExpRelax}, 
$\beta=1.0$ (normal exponential relaxation), 
and $\beta=1.3$ (compressed-exponential relaxation).
In addition, we used an STZ model 
without the rejuvenation and relaxation; the energy barrier was fixed
to $\Delta Q^{(m)}[t(\vc{x})] = k_B T \ln (1 + \delta \eta/\eta_0)$.
This STZ model show neither weakening at any deformation event, 
nor temporal variation representing the subsequent relaxation.
We refer to it as the {\it time-independent} STZ model.

Various strain rates ($\dot{\varepsilon} = \SI{e-5}{s^{-1}}$--$\SI{e+5}{s^{-1}}$) and temperatures ($T = 300$--$500$ K) were used. 
We calculated the flow stress  $\sigma_\mathrm{f}$ at each strain rate as the mean value of $\sigma_{xx}(\varepsilon)$ in the range $\varepsilon_\mathrm{s} < \varepsilon (= \dot{\varepsilon} t') \le \varepsilon_\mathrm{e} $:
\begin{align}
\sigma_\mathrm{f} 
= \frac{1}{\varepsilon_\mathrm{e} - \varepsilon_\mathrm{s}} 
\int^{\varepsilon_\mathrm{e}}_{\varepsilon_\mathrm{s}} \sigma_{xx}(\varepsilon) \intd{\varepsilon},
\end{align}
where $\sigma_{xx}(\varepsilon)$ is the global stress, i.e., the spatial average of the local stress at time $t' = \varepsilon/\dot{\varepsilon}$: $\sigma_{xx} = \ave{ \sigma_{xx}(\vc{x}) }_{\vc{x}}$, and $\varepsilon_\mathrm{s} = 0.06$ and $\varepsilon_\mathrm{e} = 0.1$ were employed.
Note that the sequence of $\sigma_{xx}(\dot{\varepsilon} t')$ is obtained only at the instants immediately before and after each deformation event, since the kMC simulation is event-driven.
Hence, a naive average of these irregular samples yields a biased estimation of $\sigma_\mathrm{f}$. To mitigate the resulting bias, we linearly interpolated $\sigma_{xx}(\dot{\varepsilon} t')$ between events and sampled the stress value at uniform time intervals before averaging.

\section{Simulation Results}
\label{sec:results}

\begin{figure}[tbp]
\centering
 \includegraphics[width=8.5cm]{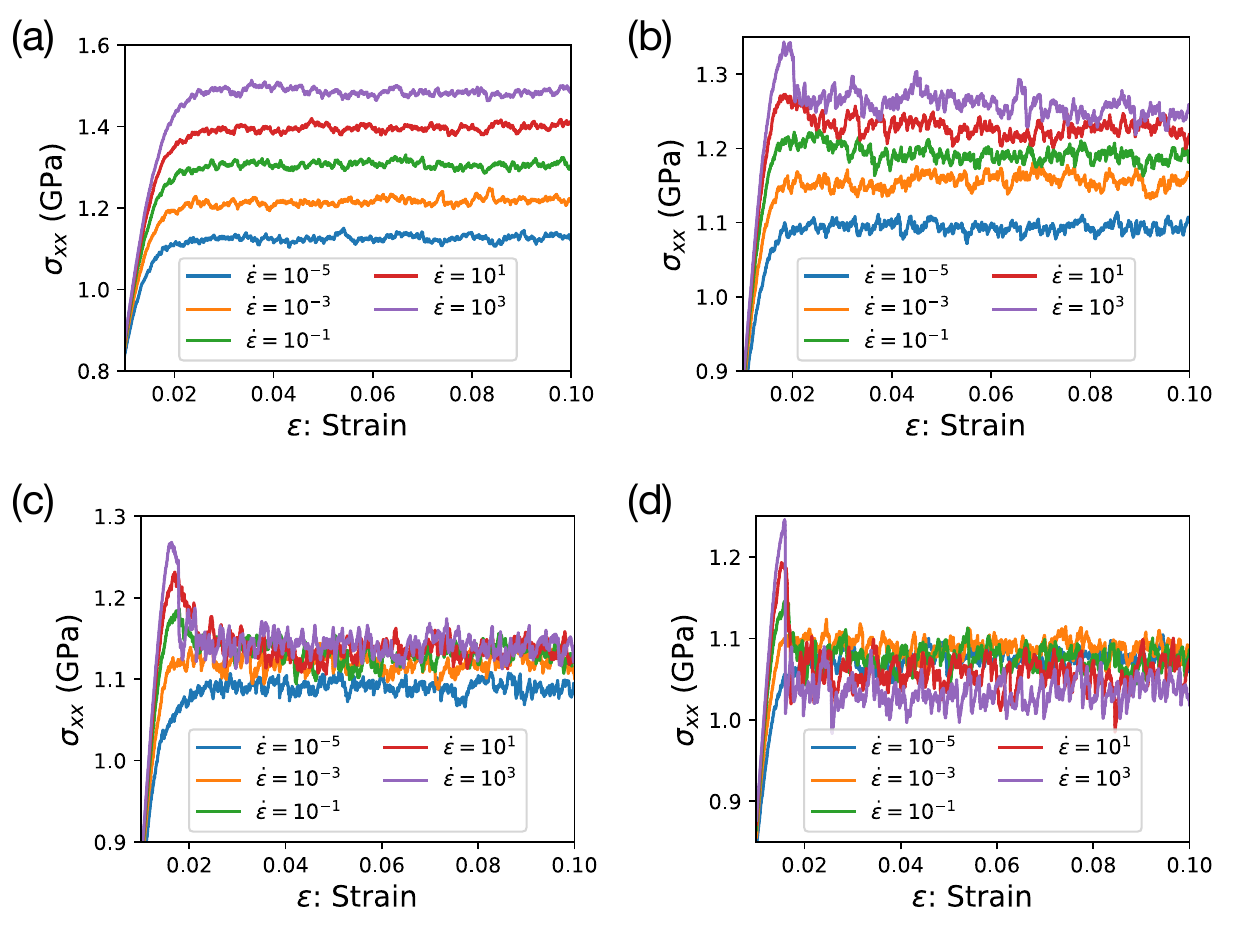}
 \caption{
Stress-strain curves at $400$~K and strain rates from $10^{-5}$ 
to $\SI{e+3}{s^{-1}}$. 
(a) Time-independent STZ model. 
(b) $\beta=0.6$. (c) $\beta = 1.0$. (d) $\beta = 1.3$.}
 \label{stst-curve}
\end{figure}

\begin{figure}[tbp]
\centering
\includegraphics[bb=6 12 521 449,width=8cm]{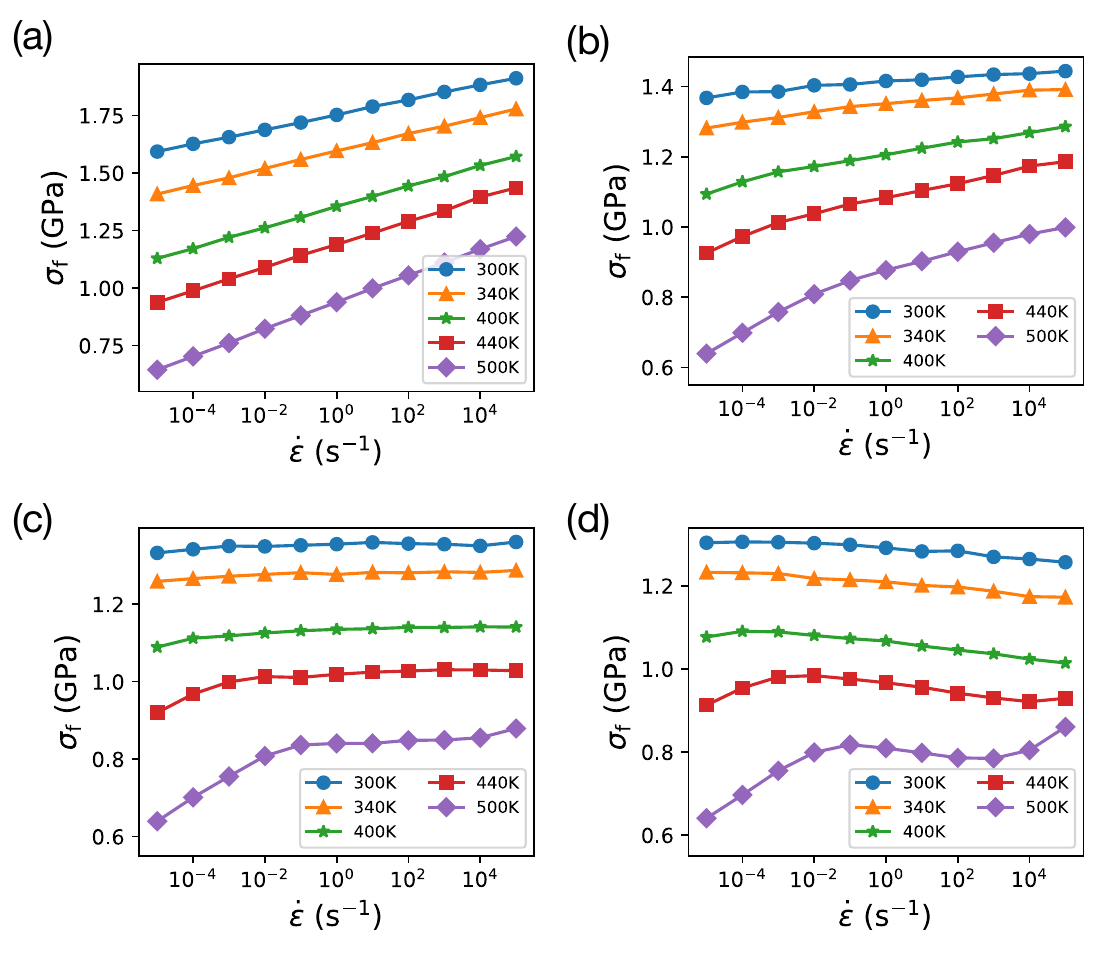}
 \caption{
Flow stress $\sigma_\mathrm{f}$ as a function of 
the strain rate $\dot{\varepsilon}$ at $T=300$, $340$, $400$, $440$, and $500$~K.
(a) Time-independent STZ model. 
(b) $\beta=0.6$. (c) $\beta = 1.0$. (d) $\beta = 1.3$.}
 \label{ratevsstress}
\end{figure}

In Fig.~\ref{stst-curve}, we show the stress-strain curves obtained from
the present kMC simulations at $400$~K and strain rates from
$\dot{\varepsilon} = 10^{-5}$ to $\SI{e+3}{s^{-1}}$ for each $\beta$ value.
Figs.~\ref{stst-curve}(a) and (b) clearly show that the flow stresses of the time-independent STZ model and the time-dependent STZ model with $\beta = 0.6$ increase with an increase in the strain rate.
On the other hand, this trend disappears at $\beta = 1.0$ and is reversed at $\beta = 1.3$ [Figs.~\ref{stst-curve}(c) and (d)].
In particular, the curve at $\dot{\varepsilon} = \SI{e+3}{s^{-1}}$ deserves attention.
Accompanying this reversal, stress serrations and the yield drops around $\varepsilon \simeq 0.03$ become prominent.
Such serrations are known to result from avalanches of STZ transformations~\cite{Bocquet2009KineticsAmorphous}.
Consistently, equivalent avalanche features are observed in the present kMC simulations, indicating that the STZ transformations are not independent events [see also Appendix~\ref{sec:avalanche}].

To confirm the strain-rate dependence observed in the stress-strain curves more systematically, we show the flow stress, $\sigma_\mathrm{f}$, as a function of strain rate $\dot{\varepsilon}$.
As seen in Fig.~\ref{ratevsstress}(a) and (b),
the time-independent STZ model and time-dependent STZ model with $\beta = 0.6$ exhibit positive SRS, i.e., an increase in flow stress with increasing strain rate, at all temperatures.
Logarithmic growth is observed over the entire regime of the time-independent STZ model [Fig.~\ref{ratevsstress}(a)] and in the high-temperature, low-strain-rate regime of the time-dependent STZ models [Figs.~\ref{ratevsstress}(b)--(d)].
For $\beta = 1.0$, the SRS is nearly zero in a large part of the regime, as can be seen in Fig.~\ref{ratevsstress}(c).
For $\beta = 1.3$, negative SRS appears over a specific range of strain rates, and the range tends to expand as the temperature decreases [Fig.~\ref{ratevsstress}(d)].

\begin{figure}[tbp]
\centering
 \includegraphics[width=8.5cm]{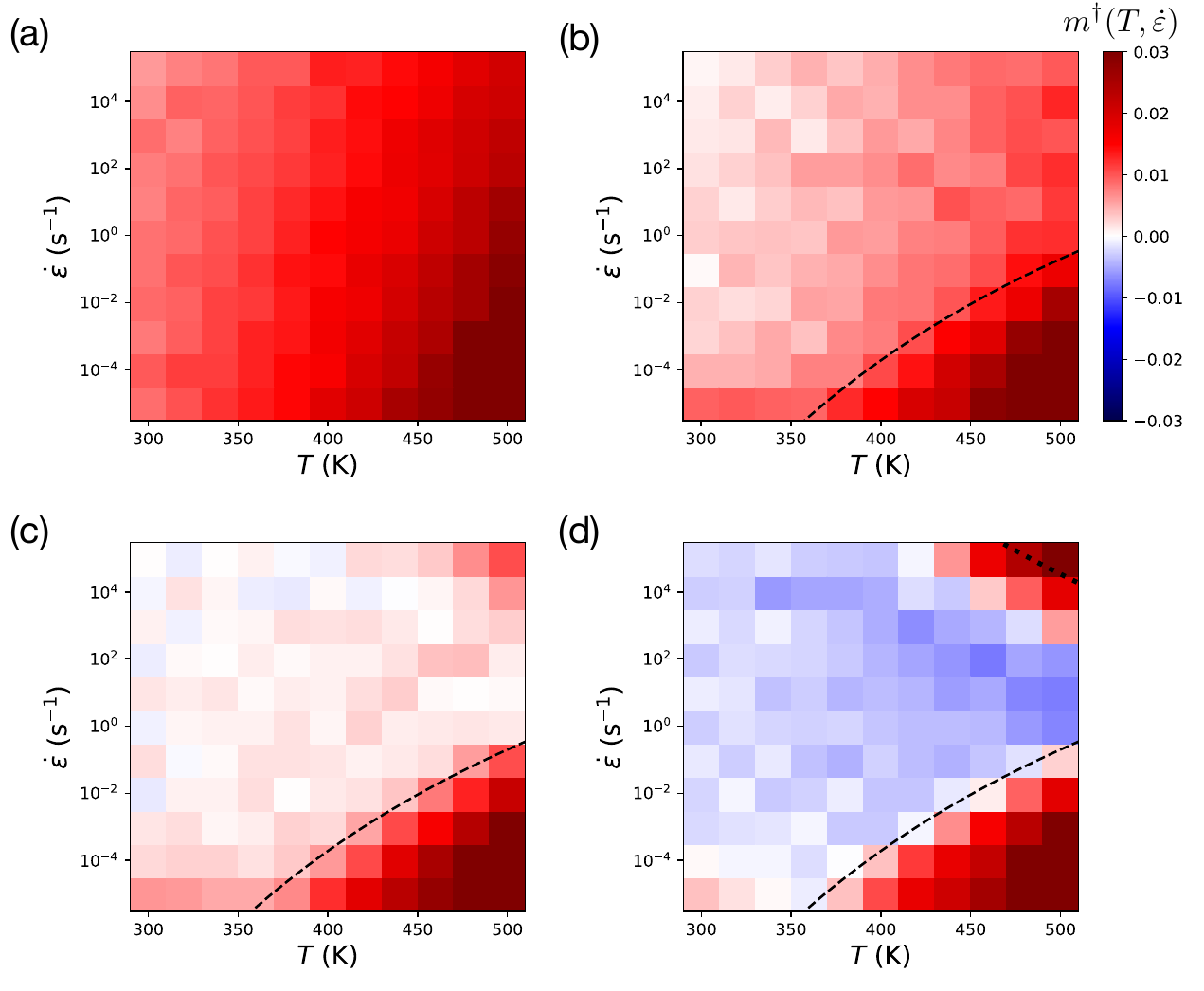}
 \caption{
Strain-rate sensitivity index, $m^\dagger(T, \dot{\varepsilon})$, obtained from 
the kMC simulations with (a) the time-independent STZ model, (b) $\beta = 0.6$,
(c) $\beta = 1.0$, and (d) $\beta = 1.3$.
The dashed and dotted lines represent the strain rates corresponding to
$t_R$ and $t_s$, respectively [see details in Sec.~\ref{sec:discussion}].}
 \label{m-map}
\end{figure}

This trend is more clearly illustrated in the $T-\dot{\varepsilon}$ map of the SRS index $m^\dagger$ in Fig.~\ref{m-map}, where the index was calculated by applying the least-squares method to the flow-stress values at strain rates $0.2 \dot{\varepsilon}$, $0.4 \dot{\varepsilon}$, $\dot{\varepsilon}$, $2 \dot{\varepsilon}$, and $4 \dot{\varepsilon}$.
Dashed and dotted lines in the figure represent the strain rates determined from $t_R$ and $t_s$, respectively (see details in Sec.~\ref{sec:discussion}).
Figure~\ref{m-map} indicates that all the STZ models exhibit positive SRS in all conditions except for $\beta = 1.3$, while fluctuations around zero are observed in the time-dependent models with $\beta = 0.6$ and $1.0$. 
In general, $m^\dagger$ tends to be large in the high-temperature and low-strain-rate regions; however, for the time-dependent models with $\beta = 1.0$ and $1.3$, we find a specific subregion at high strain rates where $m^\dagger$ becomes large.
As shown in Fig.~\ref{m-map}(d),
the negative SRS region (colored blue) appears only in the time-dependent model with $\beta = 1.3$, which exhibits compressed-exponential relaxation, and the negative region around $\dot{\varepsilon} = 10^0 \ \SI{}{s^{-1}}$ expands as the temperature decreases.

\begin{figure}[tbp]
 \begin{center}
\includegraphics[width=8cm]{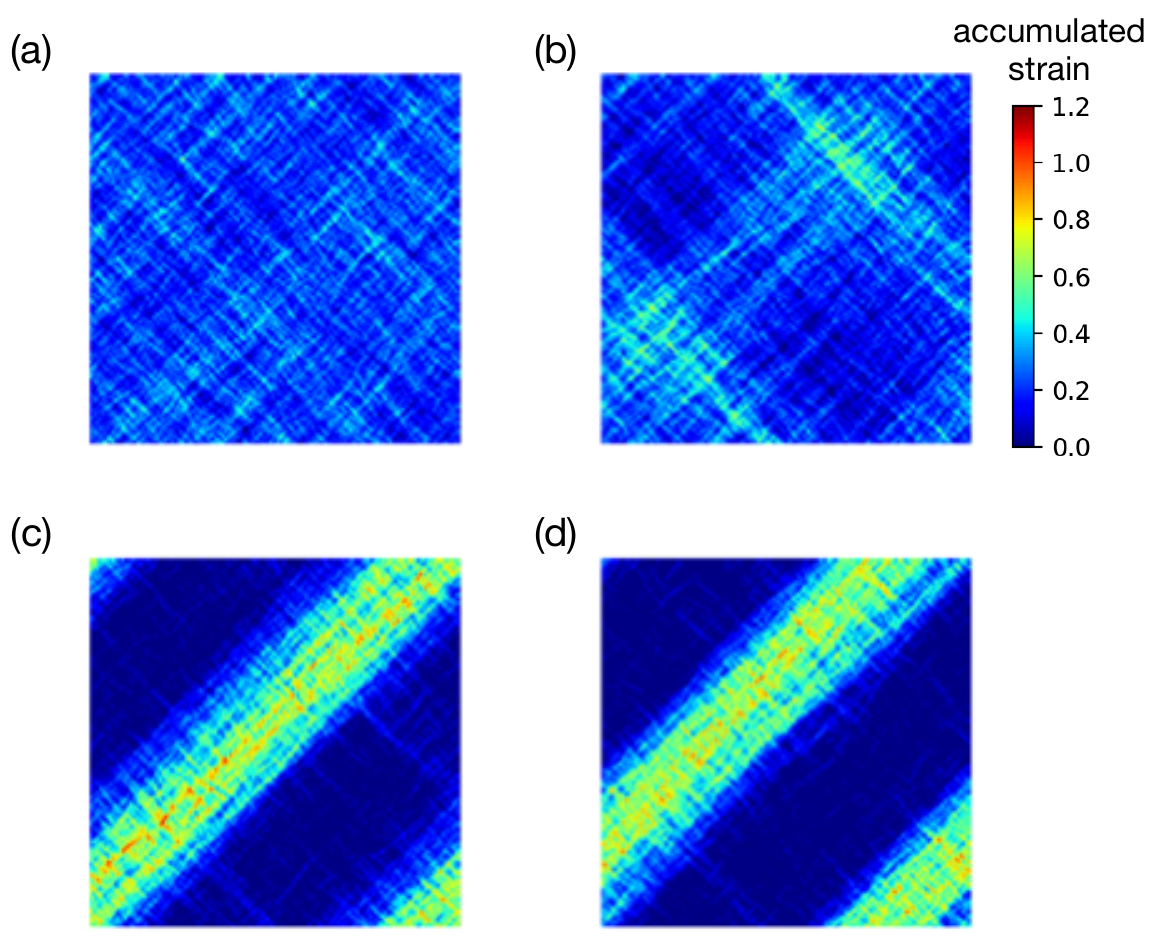}
 \caption{
The accumulated transformation strain (von Mises strain) distribution at a strain of $0.1$ at $500~\si{K}$ for the case $\beta=1.3$, shown for different strain rates:
(a) $\dot{\varepsilon} = \SI{e-4}{s^{-1}}$,
(b) $\dot{\varepsilon} = 10^0~\SI{}{s^{-1}}$,
(c) $\dot{\varepsilon} = \SI{e+2}{s^{-1}}$,
and (d) $\dot{\varepsilon} = \SI{e+5}{s^{-1}}$.}
 \label{tst_map}
 \end{center}
\end{figure}

To investigate the relationship between the negative SRS 
and the morphology of plastic deformation in metallic glasses, 
some typical snapshots of the simulation results,
including both positive and negative SRS cases, are shown in Fig.~\ref{tst_map}
[see also Fig.~\ref{ratevsstress}(d)],
where the simulation condition is $T = 500~\si{K}$ with $\beta=1.3$.
The color of the snapshots indicates the accumulated eigenstrain:
$
\tilde{\epsilon}(\vc{x}) = 
\{ \epsilon_{xx}(\vc{x})^2 + \epsilon_{yy}(\vc{x})^2 + \epsilon_{xy}(\vc{x})^2
\}^{1/2} 
$.
As shown in Fig.~\ref{tst_map}(a), the strain distribution obtained under the condition ($\dot{\varepsilon} = \SI{e-4}{s^{-1}}$),
which exhibits positive SRS, is spatially homogeneous.
In contrast, Fig.~\ref{tst_map}(c), obtained under the condition
$T = \SI{500}{K}$ with $\dot{\varepsilon} = \SI{e+2}{s^{-1}}$,
where negative SRS is observed,
indicates an apparent band-like shear localization.
Similar morphological changes have also been observed experimentally~\cite{Mukai2002,Lu2003nSRS}.
Hence, one may consider that the shear localization causes the negative SRS;
however, it should be noted that the localization remains even when
the SRS becomes positive under the condition 
$\dot{\varepsilon} = \SI{e+5}{s^{-1}}$ [Fig.~\ref{tst_map}(d)].
Further, this localization is stronger than that observed at $\dot{\varepsilon} = 10^{0}\ \SI{}{s^{-1}}$, which exhibits negative SRS [Fig.~\ref{tst_map}(b)].
This discrepancy implies that the localization is not the main factor causing the SRS but rather a consequence of it.

\begin{figure}[tbp]
\centering
 \includegraphics[width=8cm]{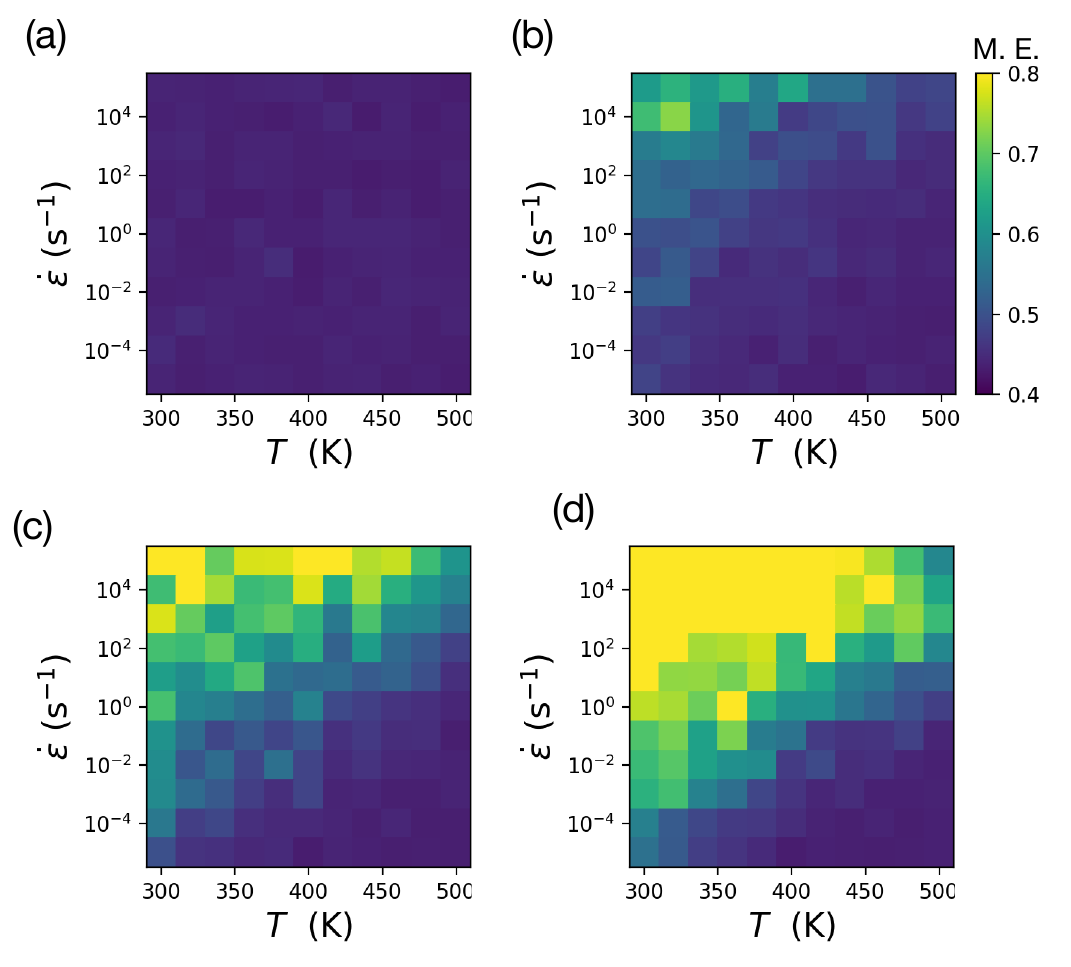}
 \caption{
The mean exceedance of the accumulated transformation strain at a strain of $0.1$.
(a) Time-independent STZ model.
(b) $\beta = 0.6$.
(c) $\beta = 1.0$.
(d) $\beta = 1.3$.}
 \label{mean-exceedance-map}
\end{figure}

To clarify the causal relationship between the localization of deformation and the negative SRS, we quantify the extent of the localization by the mean exceedance of the accumulated eigenstrain $\ave{\tilde{\epsilon}(\vc{x}')}_{\vc{x}'}$, where the mean exceedance in this study is defined as the mean value of accumulated eigenstrains at the sites $\vc{x}'$ whose accumulated strains exceed a threshold value, $\tilde{\epsilon}(\vc{x}') \ge 0.4$, at $\varepsilon = \dot{\varepsilon} t' = 0.1$.
Even though the same external strain $\dot{\varepsilon} t'$ is applied to the system, the eigenstrain (plastic strain) concentrates at a small fraction of sites when deformation is localized; consequently, the site-level accumulated strain becomes bimodal, with some sites remaining small (e.g., $\le 0.4$) and others attaining large values.
In contrast, when deformation is homogeneous, the eigenstrain is distributed uniformly across all the sites, so the sites with exceptionally large eigenstrain rarely appear.
Therefore, with an appropriately chosen threshold, the mean exceedance is expected to be large when deformation is localized and small when deformation is uniform.
In our kMC simulations, we find that certain sites undergo repeated transformations, while about half of the sites remain undeformed throughout the entire simulation run [Appendix~\ref{sec:avalanche}].
Figure~\ref{mean-exceedance-map} shows a color map of the mean exceedance as a function of $T$ and $\dot{\varepsilon}$.
The localization regime, i.e., the condition exhibiting large mean-exceedance (colored yellow), observed at $\dot{\varepsilon} = \SI{e+0}{s^{-1}}$ and $T = \SI{500}{K}$ [Fig.~\ref{mean-exceedance-map}(d)], slightly deviates from the negative SRS regime shown in Fig.~\ref{m-map}(d).
This result also suggests that the origin of the negative SRS likely lies in mechanisms other than deformation localization. This point will be discussed in the next section.

\section{Theoretical Analysis}
\label{sec:theo-analysis}

The simulation results shown in the previous section revealed that negative SRS occurs in the time-dependent STZ model with $\beta=1.3$ under the specific conditions of temperature and strain rate shown in Fig.~\ref{m-map}(d), and the contribution of deformation localization to the negative SRS is limited.
This leads to the hypothesis that strain rate, temperature, and the time dependence of the relaxation are the main factors causing the negative SRS.
Thus, we seek to understand the SRS by separating the spatial factors, e.g., deformation localization, from the temporal factors.
For this, we construct a simple theoretical model, omitting the spatial factors and the interaction between STZs, and verify whether the model reproduces the negative SRS using only time-dependent factors.

\subsection{Construction of theoretical model}

We introduce a simplified theoretical model that omits the stress field, $\sigma_{ij}(\vc{x})$, eigen-strain field, $\epsilon_{ij}(\vc{x})$, and other spatial structures by replacing them with mean field values.
For the model, we consider the steady-flow state as the balanced state between stress accumulation due to the external load $E \dot{\varepsilon} t'$ and its release through the plastic deformation (STZ transformation).
The plastic deformation of metallic glasses proceeds through the avalanche-like transformation of STZs~\cite{Maloney2004AmorphousAvalanches} (see also Appendix~\ref{sec:avalanche}), with the transformation concentrated on limited sites as shown in the simulation results.
Thus, in our model, we consider the steady-flow state to proceed through repeated transformation events at $N_d$ specific sites in the system. We further assume that the transformation of the sites is initiated when any single STZ undergoes a thermally activated transformation, leading to avalanche plasticity.
As an important simplification, we also assume that the iteration of the deformation event occurs within an interval $\bar{t}$, where the deformation event is a single collective transformation of $N_d$ STZs.

Since the global stress increases linearly by $E \dot{\varepsilon} \bar{t}$ due to the external load with strain rate $\dot{\varepsilon}$, during the time interval $\bar{t}$, the mean-field-like stress along the loading direction ($x$ direction) acting on all STZs is approximately represented by
\begin{align}
\bar{\sigma} = \sigma_0 + E \dot{\varepsilon} \bar{t},
\end{align}
where $\sigma_0$ is the mean stress when the previous deformation event is finished.
Subsequently, the transformation of the STZs releases stress as follows:
\begin{align}
\Delta \sigma = E \rho \Delta \bar{\epsilon}, 
\end{align} 
where  $\rho$ is the number density of transforming STZs $N_d/(N_x N_y)$
and $\Delta \bar{\epsilon}$ is the mean value of the transformation strain $\Delta \epsilon^{(m)}_{xx}(\vc{x})$ averaged over the position and deformation modes.
Therefore, the steady-flow state condition is described by
\begin{align}
 \dot{\varepsilon} \bar{t} = \rho \Delta \bar{\epsilon}.
\label{eq:equiv_cond}
\end{align}
The stress $\sigma_0$ satisfying the steady-flow condition of Eq.~(\ref{eq:equiv_cond}) serves as a representative measure of steady flow and can be used to characterize how the flow stress $\sigma_\mathrm{f}$ varies with strain rate.

To evaluate the mean interval $\bar{t}$, we construct the probability $P(t) \Delta t$ that a subset of STZs is activated during a short-time interval from $t$ to $t + \Delta t$.
$P(t) \Delta t$ can be considered as the probability that no STZ has activated up to $t=(l-1) \Delta t$ and any one of the STZs transforms in mode $m$ at $t=l \Delta t$, where time is discretized at intervals of $\Delta t$.
Hence, the probability is described by
\begin{align}
 P(t) \Delta t &= 
\prod_{n=1}^{N_d} \prod_{m=1}^{M} \prod_{l'=0}^{l-1} 
\left\{1 - k_n^{(m)}(l' \Delta t) \Delta t \right\}
\sum_{n'=1}^{N_d} \sum_{m'=1}^{M} k_{n'}^{(m')}(l \Delta t) \Delta t,
\end{align}
where $k_n^{(m)}(t)$ is the transformation rate of the $n$-th STZ with mode $m$.
Employing the approximation $ 1 - k_n^{(m)}(t) \Delta t \approx \exp \left[ - k_n^{(m)}(t) \Delta t \right]$, we can reduce the expression to 
\begin{align}
 P(t) \Delta t &= 
\exp \left[ - \sum_{n=1}^{N_d} \sum_{m=1}^{M} \sum_{l'=0}^{l-1}
 k_n^{(m)}(l' \Delta t) \Delta t \right]
\sum_{n'=1}^{N_d} \sum_{m'=1}^{M} k_{n'}^{(m')}(l \Delta t) \Delta t.
\nonumber
\end{align}
Upon taking the limit $\Delta t \to 0$, the sum $\sum_{l'=0}^{l-1} k_n^{(m)}(l' \Delta t) \Delta t$ is replaced by the integral $\int^t_0 k_n^{(m)}(t') \intd{t'}$; the probability is given by
\begin{align}
 P(t) \intd{t} =
\sum_{n'=1}^{N_d} \sum_{m'=1}^{M} k_{n'}^{(m')}(t) 
 \exp \left[ - \sum_{n=1}^{N_d} \sum_{m=1}^{M} \int^t_0 k_n^{(m)}(t') \intd{t'} \right]
\intd{t}.
\end{align}
Although the rate $k_n^{(m)}(t')$ is, in principle, position-dependent, we simplify the probability by replacing it with a representative spatially averaged rate.
This approximation gives us the expression for the probability:
\begin{align}
P(t) \intd{t} = 
\bar{k}(t) \exp \left[ - \int^t_0 \bar{k}(t') \mathrm{d}t' \right] \intd{t},
\label{eq:P(t)}
\end{align}
\begin{align}
  \bar{k}(t) = 
  \bar{k}_0 e^{- {\bar{G}(t)}/{k_B T} }
  =
  \bar{k}_0 \exp \left[ 
    - \frac{ \bar{Q}(t) - \bar{\sigma} \Omega /2}{k_B T} \right],
\label{eq:k_model}
 \end{align}
where $\bar{k}_0 = M N_d$, $\bar{Q}(t) = \bar{Q}_0 + \Delta Q(t)$, and $\bar{Q}_0$ is the mean of $Q_0$.
The activation-free volume $\Omega$ is estimated as $\Omega = \Delta \bar{\epsilon} V_\mathrm{STZ}$, where $V_\mathrm{STZ}$ is the volume of an STZ.
This expression indicates that our model represents a probabilistic dynamical process of a single typical STZ within the mean stress field.
Thus, the model can be considered as an effective one-body model.

Considering the probability as the waiting time distribution of the deformation event, we can evaluate the mean interval of the event by
\begin{align}
 \bar{t} = \int^{t_c}_0 t P(t) \intd{t},
\label{eq:t_integral}
\end{align}
where $t_c$ is the time when the activation barrier $G(t)$ is zero.
We emphasize that the mean interval can be represented as a function of the typical stress, $\bar{t} = \bar{t}(\sigma_0)$, since $\bar{k}(t)$ depends on the mean stress as shown in Eq.~(\ref{eq:k_model}).
Finding $\sigma_0$ satisfying the steady-flow condition represented by Eq.~(\ref{eq:equiv_cond}) using Eq.~(\ref{eq:t_integral}), we can obtain the typical stress representing the flow stress. 
Thus, obtaining the explicit representation of $\sigma_0$ is the main aim of the present theoretical analysis.
However, the integral of Eq.~(\ref{eq:t_integral}) is so complicated that Eqs.~(\ref{eq:equiv_cond}) and (\ref{eq:t_integral}) cannot be solved for $\sigma_0$ in closed form.

Here we present the numerical solution of the theoretical model, which was obtained by the following procedure.
We integrated Eq.~(\ref{eq:t_integral}) using the trapezoidal method and found $\sigma_0$ using the bisection method.
The numerical integration was carried out from $t=\SI{e-20}{s}$ to $t_c$ using $2000$ logarithmically spaced subintervals, and $t_c$ was obtained by bisection.
For the evaluation, we employed the parameter values $\Delta \bar{\epsilon} = 0.1$ and $\rho = 1/4$, which were estimated in the kMC simulations (see Appendices~\ref{sec:plast-strain} and \ref{sec:avalanche}).
Further, the other parameters were set to the same values as those used in the simulations.

\begin{figure}[tbp]
 \centering
\includegraphics[bb=6 12 521 221,width=8cm]{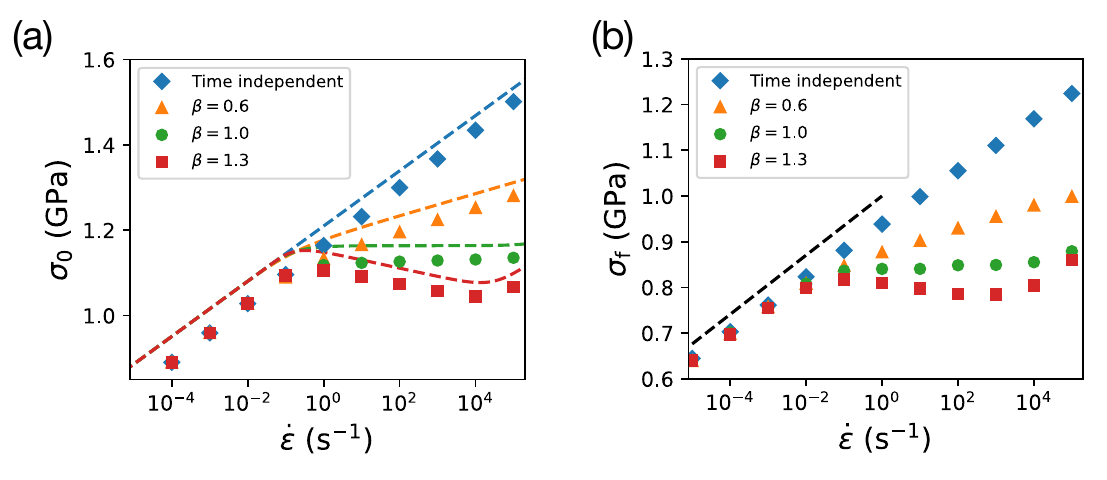}
\caption{
The representative stress in the steady-flow state as a function of strain rate $\dot{\varepsilon}$ at $\SI{500}{K}$.
(a) The representative stress that is numerically calculated based on the simplified theoretical model. The dashed lines show approximate solutions.
(b) The flow stress obtained by the kMC simulations.
}
\label{fig:sgm_0-num_sol}
\end{figure}

The numerical values of $\sigma_0$ at $T=\SI{500}{K}$ based on our theoretical model are shown in Fig.~\ref{fig:sgm_0-num_sol}(a),
while Fig.~\ref{fig:sgm_0-num_sol}(b) indicates the flow stress $\sigma_\mathrm{f}$ resulting from the kMC simulations, where the blue diamonds represent the time-independent STZ model, and orange triangles, green circles, and red squares represent the time-dependent STZ models with $\beta = 0.6, 1.0$, and $1.3$, respectively.
We can confirm quantitative agreement between the numerical solutions and the kMC simulation results.
Although the flow stress is overestimated in the results of our theoretical model, it is noteworthy that the strain-rate regime associated with the negative SRS closely coincides with that of kMC simulations.
Therefore, we conclude that the present model captures the essential behavior of the negative strain-rate dependence obtained in our kMC simulations.

\subsection{Approximate solution of theoretical model}
\label{sec:sol_approx}

The numerical evaluation presented in the previous subsection supports the validity of our theoretical model that exhibits negative SRS.
By introducing several approximations into the model, we obtain an approximate closed-form solution and analyze the underlying mechanism responsible for negative SRS.

Instead of solving Eq.~(\ref{eq:equiv_cond}) for $\sigma_0$, replacing $\bar{t}$ by the peak time $t^*$ of the waiting time distribution $P(t)$, we deduce $\sigma_0$ approximately.
The distribution described by Eq.~(\ref{eq:P(t)}) is expected to have a sharp peak at a certain point $t^*$, since the distribution is composed of two exponential functions.
Thus, the mean time $\bar{t}$ should be around $t^*$, where $t^*$ is the time satisfying the condition $\pdev{P}{t}{} = 0$, that is,
\begin{align}
\left. \dev{\bar{k}}{t}{} \right|_{t=t^*}  - \bar{k}^2(t^*) = 0.
\end{align}
While the solution of the equation for $t^*$ is still difficult to obtain, by substituting Eq.~(\ref{eq:k_model}) into the equation, we obtain the expression:
\begin{align}
- \frac{\dot{Q}(t^*) - E \Omega \dot{\varepsilon}/2 }{k_B T} 
= \bar{k}_0 \exp \left[ - \frac{ Q(t^*) 
- (E \dot{\varepsilon} t^* + \sigma_0)\Omega/2 }{k_B T} \right].
\end{align}
This allows us to solve the equation for $\sigma_0$ as:
\begin{align}
 \sigma_0 = 
  \frac{2 k_B T}{\Omega}
 \ln \left[ \frac{\Omega E \dot{\varepsilon}}{2} 
- \dot{Q}( t^* ) \right]
&+\frac{2}{\Omega}  Q( t^* ) 
\nonumber \\&
- E \rho \Delta \bar{\epsilon} - \frac{2 k_B T}{\Omega} \ln (\bar{k}_0 k_B T).
\label{eq:sigma_theo}
\end{align}
Regarding $t^*$ as the mean interval $\bar{t}$, we can apply the steady-state condition in Eq.~(\ref{eq:equiv_cond}) to the equation:
\begin{align}
t^* = \rho \Delta \bar{\epsilon} / \dot{\varepsilon}. 
\label{eq:t^*}
\end{align}
Hence, we obtain the expression for $\sigma_0$ as a function of $\dot{\varepsilon}$.

In the expression for $\sigma_0$, only the first two terms depend on the strain rate, $\dot{\varepsilon}$, while the third and fourth terms do not contain any strain-rate dependence; they serve only to offset the flow stress globally.
Only the factor $\ln \Omega E \dot{\varepsilon}$ included in the first term raises the flow stress as the strain rate increases, independent of $Q(t)$.
Hence, the flow stress increases logarithmically with the strain rate in the regime where $Q(t)$ is considered to be almost constant, namely, the entire regime of the time-independent STZ model and the pre-relaxation and post-relaxation regimes of the time-dependent STZ model (see Fig.~\ref{fig:Q_t}).
Indeed, the logarithmic growth appears in the entire regime of the time-independent STZ model and the slow strain-rate regime, $\dot{\varepsilon} \le \SI{e-1}{s^{-1}}$, of the time-dependent STZ model as shown in Figs.~\ref{ratevsstress} and \ref{fig:sgm_0-num_sol}, where the slow strain rate gives sufficient time for $Q(t)$ to relax and saturate (see Fig.~\ref{fig:Q_t}).
As shown in Fig.~\ref{fig:sgm_0-num_sol}(b), the kMC simulations showed that the flow stress and the increasing rate of the logarithmic growth in the regime are common to all the STZ models. This feature is explained by the fact that the first term has no model-dependent parameters except for $\dot{Q}$, which is negligible when $Q(t)$ is constant.
Applying $\dot{Q} = 0$ to Eq.~(\ref{eq:sigma_theo}), we obtain the rate of increase as:
\begin{align}
 \sigma_0/\ln \dot{\varepsilon} = 2 k_B T / \Omega.
\label{eq:sgm_theo_Q_t_const}
\end{align}
The dashed straight line in Fig.~\ref{fig:sgm_0-num_sol}(b) indicates this increasing rate, and the theoretical estimation shows remarkable agreement with the kMC simulation results.

\begin{figure}[tbp]
\centering
 \includegraphics[width=8.5cm]{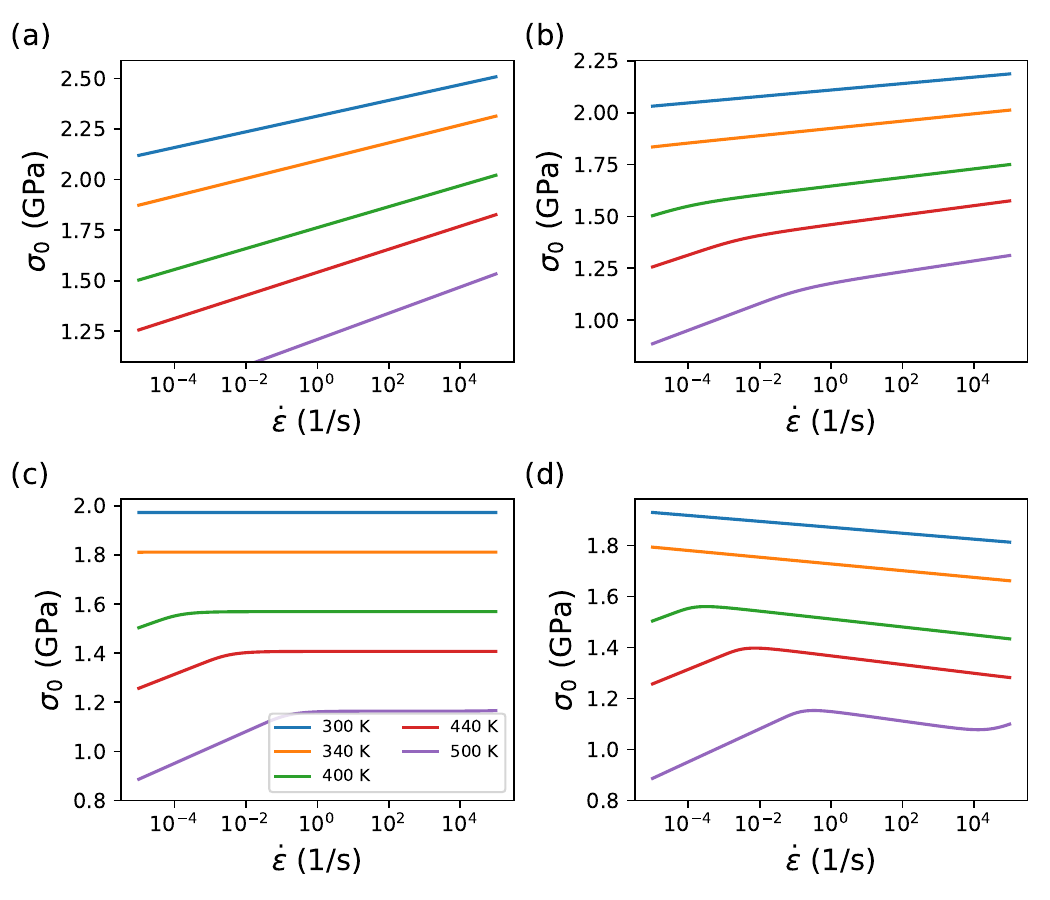}
 \caption{
The flow stress obtained by the theoretical evaluation as a function of strain rate $\dot{\varepsilon}$ on a semi-logarithmic scale at $300$, $340$, $400$, $440$, and $500$~K for 
(a) the time-independent STZ model (without rejuvenation or relaxation) and the time-dependent STZ model with (b) $\beta=0.6$, (c) $\beta =1.0$,
and (d) $\beta = 1.3$.}
 \label{fig:flowcurves-theo}
\end{figure}

The dashed curves in Fig.~\ref{fig:sgm_0-num_sol}(a) are the flow curves obtained by applying the activation energy function used in these simulations [Eqs.~(\ref{eq:Q_t}) and (\ref{eqn:deltq})] to an approximate solution described in Eq.~(\ref{eq:sigma_theo}).
Here, the time derivative of the KWW function is represented by
\begin{align}
 \dev{Q}{t}{} &=
k_B T \frac{\beta}{t_R} \left( \frac{t}{t_R} \right)^{\beta-1}
\frac{ A(t)}{1 + {\delta \eta}/{\eta_0} - A(t)},
\end{align}
where we introduced a new symbol and used its derivative for the sake of visibility:
\begin{align}
 A(t) &\equiv ({\delta \eta}/{\eta_0}) e^{-(t/t_R)^{\beta}},
\\
\dev{A}{t}{} &= -\frac{\beta}{t_R} \left( \frac{t}{t_R} \right)^{\beta-1} A(t).
\end{align}
In Fig.~\ref{fig:sgm_0-num_sol}(a),
although the theoretical estimations of $\sigma_0$ (dashed curves) and the numerical solutions (points) show some discrepancy because we replaced $t^*$ with $\bar{t}$ as the approximation, the error remains within about 5 \%.

The flow curves obtained by the theoretical estimation under the conditions corresponding to those shown in Fig.~\ref{ratevsstress} are depicted in Fig.~\ref{fig:flowcurves-theo}. The figure also indicates significant qualitative agreement between the theoretical and simulation results.

The above discussion supports the validity of our theoretical model and the approximate solution described in Eq.~(\ref{eq:sigma_theo}).
It should be noted that our theoretical model and its approximate solution are not restricted to the KWW-type activation energy considered here; the model is generally applicable to materials whose deformation is governed by thermal activation.

\subsection{Strain rate sensitivity}
\label{sec:srs-theo}

Based on the approximate solution obtained in the previous subsection [Eq.~(\ref{eq:sigma_theo})], we derive the SRS index $m^\dagger$.
Using the derivative of the solution
\begin{align}
 \del{\ln \sigma_0}{\ln \dot{\varepsilon}}{}
= 
 \frac{\dot{\varepsilon}}{\sigma_0} \del{\sigma_0}{\dot{\varepsilon}}{},
\end{align}
we obtain the following expression for the index:
\begin{align}
 m^\dagger = \frac{\dot{\varepsilon}}{\sigma_0} \frac{2 }{\Omega}
\left\{
k_B T \frac{\Omega E/2 
- \pdev{\dot{Q}(t^*)}{\dot{\varepsilon}}{} }
{\Omega E \dot{\epsilon} /2 - \dot{Q}(t^*) }
+ \dev{ Q( t^* ) }{\dot{\varepsilon}}{} 
\right\},
\label{eq:m_theo}
\end{align}
where
\begin{align}
 \dev{Q(t^*)}{\dot{\varepsilon}}{} = 
- \frac{ \rho \Delta \bar{\epsilon} }{\dot{\varepsilon}^2} 
\dev{Q(t^*)}{{t^*}}{}, \quad
 \dev{\dot{Q}(t^*)}{\dot{\varepsilon}}{} = 
- \frac{ \rho \Delta \bar{\epsilon} }{\dot{\varepsilon}^2} 
\dev{Q(t^*)}{{t^*}}{2}.
\end{align}

\begin{figure}[tbp]
 \begin{center}
 \includegraphics[width=8.5cm]{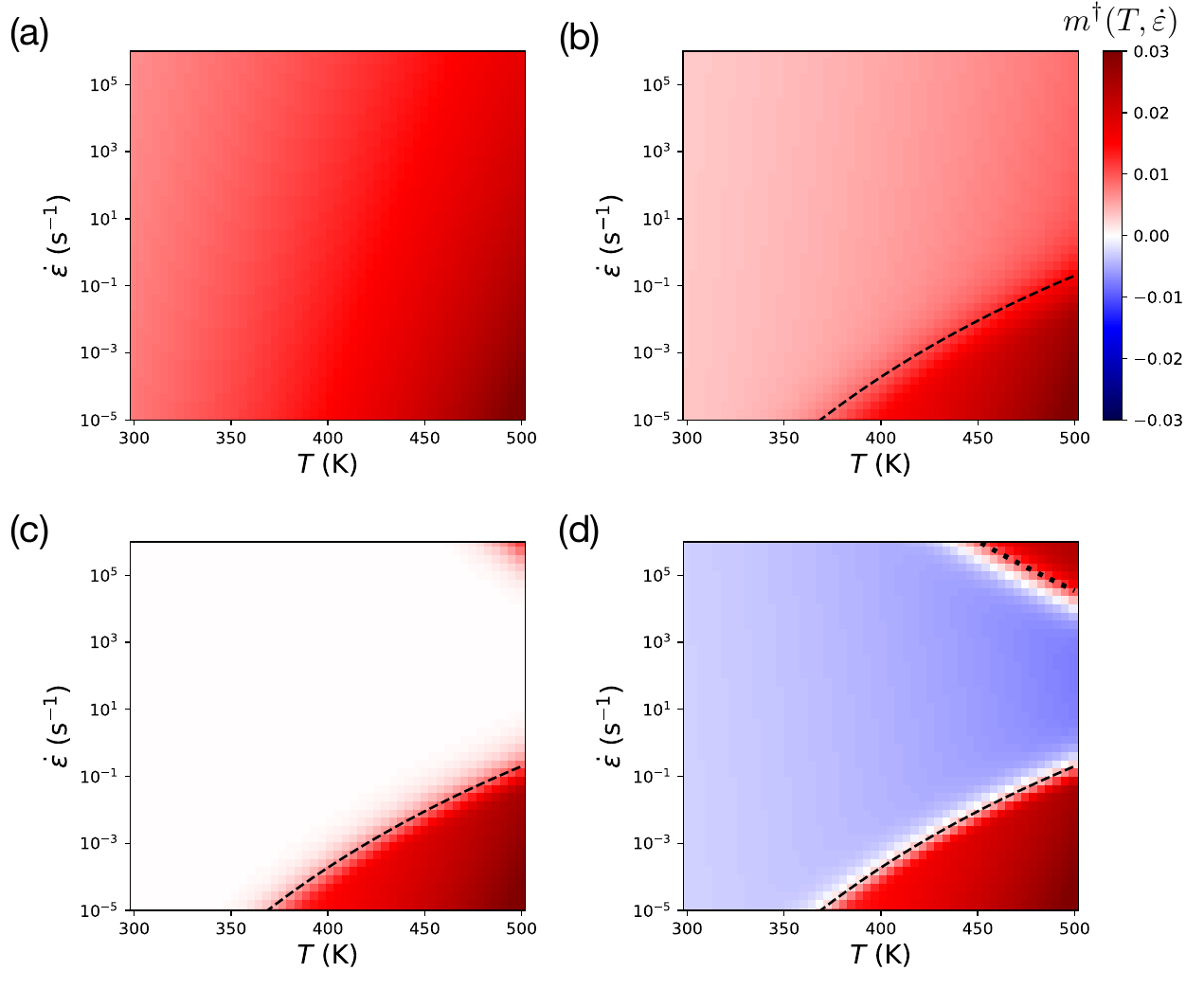}
 \caption{
Strain-rate sensitivity map $m^\dagger(T, \dot{\varepsilon})$ obtained from
the theoretical evaluation for (a) the time-independent model (no rejuvenation nor relaxation) and the time-dependent model with (b) $\beta=0.6$, (c) 1.0 and (d) 1.3. 
The dashed and dotted lines represent the strain rates corresponding to $t_R$ and $t_s$, respectively [see Sec. V for details].}
 \label{m-map-theo}
 \end{center}
\end{figure}

We show color maps of the index $m^\dagger(T, \dot{\varepsilon})$ for the activation energies used in this study as a function of temperature and strain rate in Fig.~\ref{m-map-theo}, where the second derivative of the activation energy is 
\begin{align}
 \dev{Q}{t}{2} = 
- \dev{Q}{t}{} \frac{\beta}{t} 
\left[
\frac{\beta-1}{\beta} 
- \left( \frac{t}{t_R} \right)^{\beta} 
\frac{ 1 + {\delta \eta}/{\eta_0} }{1 + {\delta \eta}/{\eta_0} - A(t) }
\right].
\end{align}
The dependence of $m^\dagger$ on strain rate and temperature shown in the figure is clearly consistent with that obtained from the kMC simulations shown in Fig.~\ref{m-map}.

\section{Discussion}
\label{sec:discussion}

We derived the expressions for the typical flow stress $\sigma_0$ and the SRS index $m^\dagger$ from our simplified theoretical model as shown in Eqs.~(\ref{eq:sigma_theo}) and (\ref{eq:m_theo}), respectively; however, the mechanism responsible for the negative SRS, namely the role of $\beta$, remains unclear due to the complexity of the expressions.
In this section, we reveal the mechanism and the role of $\beta$ by applying some approximations to the theoretical expressions.

Because the complexity of the solutions comes from the form of the KWW-type activation-energy function $Q(t)$, we replace it with the approximate representation in the transition regime depicted in Eq.~(\ref{eq:Q_approx-trans}).
This replacement simplifies the time derivative of $Q(t)$ to tractable forms:
\begin{align}
 \dev{Q}{t}{} = \frac{\beta k_B T}{t}, \quad
 \dev{Q}{t}{2} = -\frac{\beta k_B T}{t^2}.
\label{eq:dQdt-KWW-approx}
\end{align}
Using the approximate expressions, we describe the flow stress as
\begin{align}
\sigma_0 =& 
   \frac{2 k_B T}{\Omega} \left( 1 - \beta \right) \ln \dot{\varepsilon} 
- E \rho \Delta \bar{\epsilon} 
\nonumber \\&
+\frac{2 k_B T}{\Omega} \left\{ 
 \ln \left[ \frac{1}{\bar{k}_0} \frac{\delta \eta}{\eta_0} \left(
\frac{\Omega E}{2  k_B T} 
- \frac{1}{ \beta \rho \Delta \bar{\epsilon} } \right) \right] 
+ \frac{Q_0}{k_B T} + \beta \ln \rho \Delta \bar{\epsilon}  \right\}.
\label{eq:sigma_theo_KWW-approx}
\end{align}
Now we can find the strain-rate dependence only in the first term of the equation.
As is readily seen from the equation, the sign of SRS, i.e., the dependence of the flow stress on the strain rate, is determined by $(1-\beta)$.
This is the reason why the negative SRS emerges only for $\beta = 1.3$ in the kMC simulations.

From the discussion of the strain-rate dependence in the regime of $\dot{Q} = 0$ [Eq.~(\ref{eq:sgm_theo_Q_t_const})] and the analysis above, it is evident that the negative SRS originates from the time dependence of $Q(t)$ and appears only in the regime where $Q(t)$ varies with time.
As depicted in Fig.~\ref{fig:Q_t}, the time variation of $Q(t)$ appears in the range from $t_s$ to $t_R$ in the KWW-type relaxation.
By utilizing the steady-flow state condition of Eq.~(\ref{eq:equiv_cond}), the two time scales can be converted to characteristic strain rates: 
$\dot{\varepsilon}_s = \rho \Delta \bar{\epsilon} t_s$
and 
$\dot{\varepsilon}_R = \rho \Delta \bar{\epsilon} t_R$, respectively.
We depict the strain rates as dashed lines in Fig.~\ref{m-map-theo}, where the strain rates are represented as a function of temperature by using Eqs.~(\ref{eq:t_R}) and (\ref{eq:t_start}).
As shown in the figure, the characteristic strain rates are well aligned with the boundaries where the SRS changes abruptly, particularly those where the negative SRS arises in Fig.~\ref{m-map-theo}(d).
Furthermore, it is noteworthy that they also correspond closely to the variations in $m^\dagger$ obtained from the kMC simulation results, as shown in Fig.~\ref{m-map}.
This also corroborates the validity of our claim that the time variation of the activation energy is the main factor resulting in the negative SRS.

Finally, considering the competition between the time scale of the external loading and the STZ relaxation,  we discuss a general condition for negative SRS that is not restricted to the KWW-type activation-energy function considered above.
Let us assume that the relaxation of the activation energy of STZs is relatively slow, such that the contribution of $\dot{Q}$ in Eq.~(\ref{eq:sigma_theo}) is negligible.
By this simplification, the strain-rate sensitivity is represented by
\begin{align}
 m^\dagger &= 
\frac{1}{\sigma_0} \frac{2 }{\Omega}
\left\{ k_B T - \frac{ \rho \Delta \bar{\epsilon} }{\dot{\varepsilon}} 
\dev{Q(t^*)}{{t^*}}{}  \right\},
\end{align}
where we used the relation in Eq.~(\ref{eq:t^*}).
This equation clearly indicates that the factor causing the negative SRS is the time dependence of $Q(t^*)$, where the factor appears as the second term in Eq.~(\ref{eq:sigma_theo}) and as the second term within parentheses in Eq.~(\ref{eq:m_theo}). 

From the above representation, the negative SRS condition is given by
\begin{align}
 k_B T \frac{ \dot{\varepsilon} }{ \rho \Delta \bar{\epsilon} } 
=  \frac{ k_B T }{ t^*} 
< \dev{Q(t^*)}{{t^*}}{}.
\label{eq:cond-negative-SRS}
\end{align}
This inequality indicates that the negative SRS appears when the increasing rate of the activation energy exceeds a characteristic rate mainly determined by the product of temperature and strain rate.
One can confirm that the inequality is consistent with the discussion of the KWW-type function given in Eqs.~(\ref{eq:dQdt-KWW-approx}) and (\ref{eq:sigma_theo_KWW-approx}).
Interestingly, the inequality described by Eq.~(\ref{eq:cond-negative-SRS}) coincides with the negative-SRS condition proposed by Dubach and coworkers [Eq.~(\ref{eq:Dubach})] based on a constitutive model~\cite{Dubach2009modelSRSofBMG}, if $\Delta g$ is assumed to be $Q(t)$ and our steady-flow state condition shown in Eq.~(\ref{eq:equiv_cond}) is employed.
In other words, the present kMC simulations and our micro-scale probabilistic model support the conclusions of the work based on the constitutive model~\cite{Dubach2009modelSRSofBMG}.

Our theoretical model and kMC simulations, which are consistent with previous studies discussed above, indicate that negative strain-rate sensitivity can emerge in metallic glasses exhibiting compressed-exponential relaxation ($\beta > 1$). 
However, although negative SRS has been reported experimentally in various metallic glasses, clear evidence of compressed-exponential relaxation has only been found in a limited number of cases~\cite{Morishita2012CompressedExpRelax}. 
This discrepancy remains an open issue. One possible explanation is that rejuvenation by deformation and local heating may alter the relaxation behavior only in the deformed region, thereby making the relaxation effectively compressed-exponential relaxation in the deformation region.
Directly observing such local changes in the relaxation property is challenging, but future studies based on atomistic simulations and related approaches may clarify this point.

The good agreement between our simulation results and theoretical predictions indicates that the model has explanatory power for negative SRS in metallic glasses.
Our model excludes spatial information on deformation; hence, it suggests that deformation localization, although observed in the present simulations and reported in previous studies, does not directly contribute to the negative SRS.
Needless to say, the present result does not rule out indirect or synergistic effects of other factors, such as deformation localization and morphology, free volume, and local heating.
The contribution of each factor to SRS will be elucidated by extending the kMC simulations and the theoretical model.

\section{Summary}
In this study, we performed kinetic Monte Carlo (kMC) simulations to elucidate the mechanism responsible for negative strain-rate sensitivity (SRS) in metallic glasses.
For this, we derived a time-dependent intrinsic energy barrier for shear transformation zones (STZ) that exhibits KWW-type relaxation behavior characterized by an exponent $\beta$, and introduced some kMC simulation protocols to capture the essential features of the plastic deformation of metallic glasses.
We found that the SRS tends to decrease in a specific regime determined by strain rate and temperature, and that the reduction becomes significant, leading to negative SRS, when the KWW-type relaxation is represented by a compressed-exponential function ($\beta > 1$). 
Shear localization is observed when the SRS becomes relatively low or negative; however, the correspondence between the shear localization and the negative SRS is limited.

To explain the condition for negative SRS in the simulations, we proposed an effective one-body theoretical model that neglects spatial structures, such as distributions of local stress, transformation strain, and energy barriers.
The theoretical model reproduces SRS trends consistent with those obtained from kMC simulations, even though the spatial structures are omitted.
Further, an approximation using the maximum-term method yields an analytical expression for the characteristic stress in the steady-flow state, 
and the expression captures the strain-rate dependence of the flow stress that is consistent with the results of the present kMC simulations.
The analysis of the approximate solution explains the reason why the negative SRS is observed in the kMC simulations with the compressed-exponential function ($\beta > 1$).
The analysis also elucidates a general condition for negative SRS: an explicit inequality describing the competition between the time scale of external loading and the relaxation time scale of the energy barrier.

\begin{acknowledgements}
This study was supported by Ministry of Education, Culture, Sports, Science and Technology (MEXT) as ``Exploratory Challenge on Post-K computer'' (Challenge of Basic Science--Exploring Extremes through Multi-Physics and Multi-Scale Simulations) and Japan Society for the Promotion of Science (JSPS) KAKENHI (Grant Number 23K03259, 18K13658, 21K03771, and 23K20037).
S.O. was supported by MEXT Programs (Grant Numbers JPMXP1122684766, JPMXP1020230325, and JPMXP1020230327).
\end{acknowledgements}


\appendix

\section{Collective and iterative STZ activation}
\label{sec:avalanche}

Plastic deformation in amorphous solids, including metallic glasses, proceeds not through randomly independent activation of STZs, but through collective activation of STZs, also known as avalanches, which exhibit power-law statistics represented by $P(x) \propto x^{-\kappa}$, where $x$ and $\kappa$ are the event size and a characteristic exponent, respectively~\cite{SOC1998Jensen,Salerno2012AmorphousAvalanches,Maloney2006AmorphousAvalanche}.
We confirm that such avalanche dynamics, i.e., chain-reaction activation of multiple STZs, appear in our kMC simulations.

\begin{figure}[tbp]
 \centering
\includegraphics[bb=6 12 589 439,width=8cm]{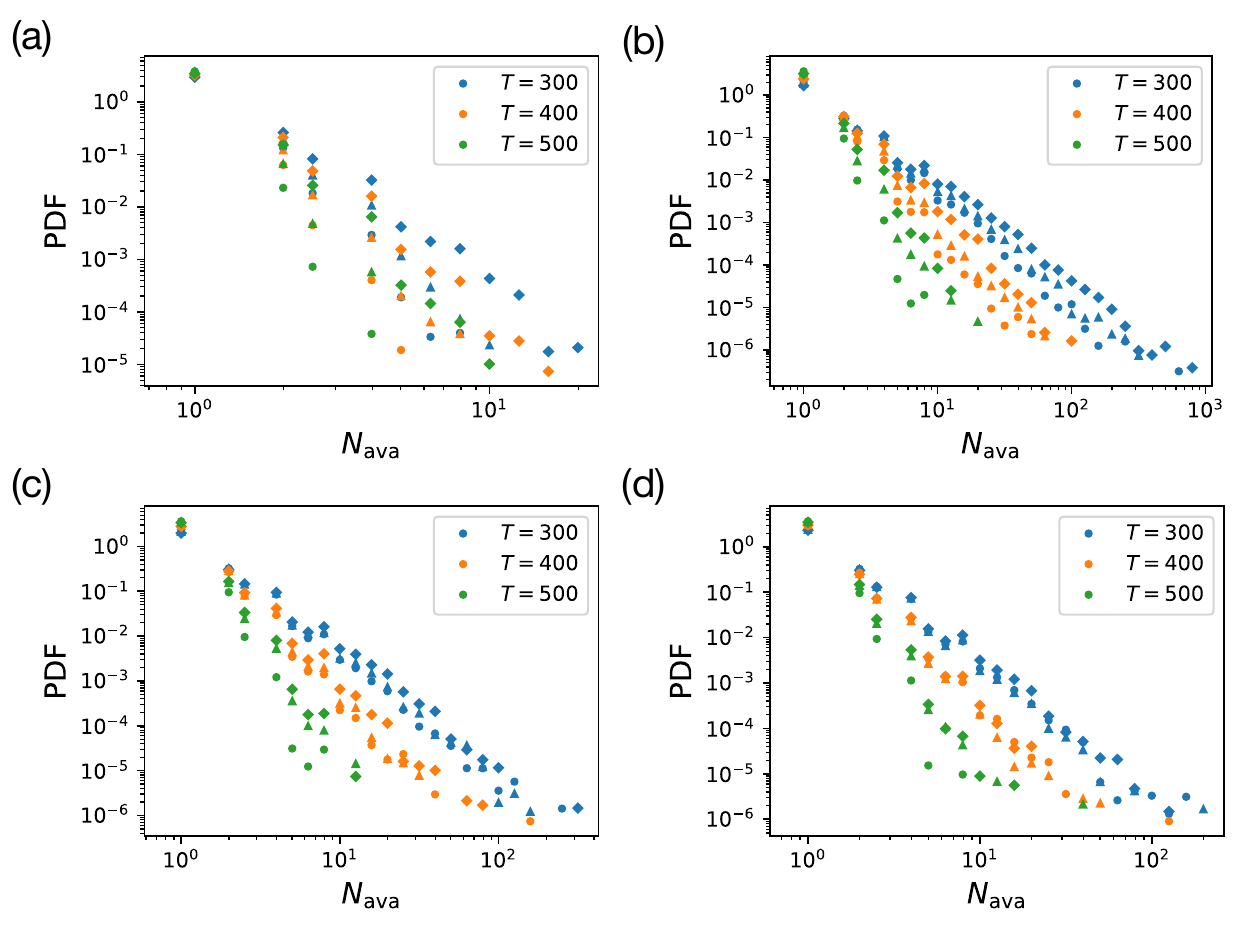}
\caption{The statistical distribution of the number of STZs activated in a single plastic deformation (avalanche) event in the kMC simulations.
(a) Time-independent STZ model, (b) $\beta = 0.6$, (c) $\beta = 1.0$,
(d) $\beta = 1.3$.}
\label{fig:dist-N_ava}
\end{figure}

We focused on deformation events that occurred in the flow regime ($0.06 \le \varepsilon \le 0.1$) and counted the number of sites that transformed during each avalanche deformation event.
The probability distribution $P(N_\mathrm{ava})$ was obtained by counting the frequency with which the number of chain-transformed sites equals $N_\mathrm{ava}$.
In Fig.~\ref{fig:dist-N_ava}, the distributions obtained under representative conditions of temperature and strain rate are shown.
As can be confirmed from the figure, all the distributions indicate algebraic decay, that is, power-law decay, although this behavior tends to be obscured under high temperature conditions.
Hence, these results validate the existence of the avalanche-like STZ transformation in the present kMC simulations for metallic glass plasticity.

\begin{figure}[bp]
 \centering
\includegraphics[bb=6 14 583 439,width=8cm]{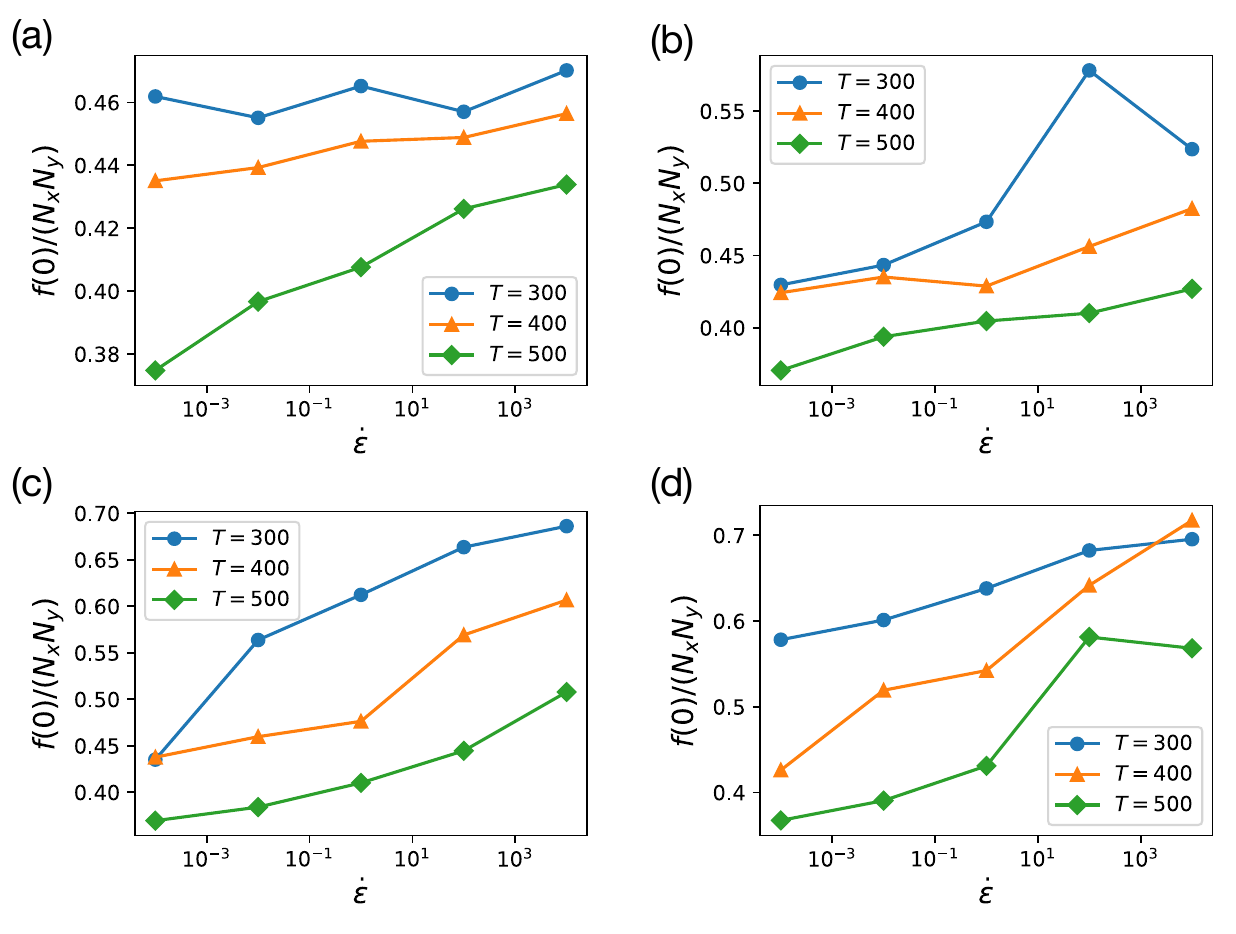}
\caption{
The ratio of undeformed sites as a function of strain rate.
(a) Time-independent STZ model, (b) $\beta = 0.6$, (c) $\beta = 1.0$,
(d) $\beta = 1.3$.}
\label{fig:ratio-N_defo}
\end{figure}

Furthermore, by counting the total number of transformations at each site,
$N_\mathrm{defo}(n_x, n_y)$, we obtained the frequency distribution $f(n_\mathrm{defo})$ of sites that undergo $n_\mathrm{defo}$ cumulative transformations:
\begin{align}
 f(n_\mathrm{defo})
 = \sum_{n_x=1}^{N_x} \sum_{n_y=1}^{N_y} \delta_{n_\mathrm{defo}, N_\mathrm{defo}(n_x, n_y)},
\end{align}
where $\delta_{i, j}$ is the Kronecker delta.
Using the frequency distribution, we calculated the ratio of undeformed sites to all the sites: $f(0)/(N_x N_y)$.
Figure~\ref{fig:ratio-N_defo} shows the ratio as a function of the strain rate.
As can be seen from the figure, although the fraction of undeformed sites decreases with increasing temperature and $\beta$, at least $\sim40\%$ of sites remain undeformed.
In addition, we confirmed that the ratio exceeds $70\%$ if the sites that deform only once are also included: $[f(0)+f(1)]/(N_x N_y) > 0.7$.
These results indicate that the deformation events occur repeatedly only within a limited subset of sites.
Although a unique value of $f(0)/(N_x N_y)$ cannot be specified since the value depends on the conditions, we used a rough representative value $\rho = 1/4$ for the theoretical analysis in Sec.~\ref{sec:theo-analysis}.
Note that the value of $\rho$ has no influence on the qualitative results of the theoretical analysis in this study.

\section{Statistical distribution of transformation strain}
\label{sec:plast-strain}

\begin{figure}[tbp]
 \centering
\includegraphics[bb=6 12 589 439,width=8cm]{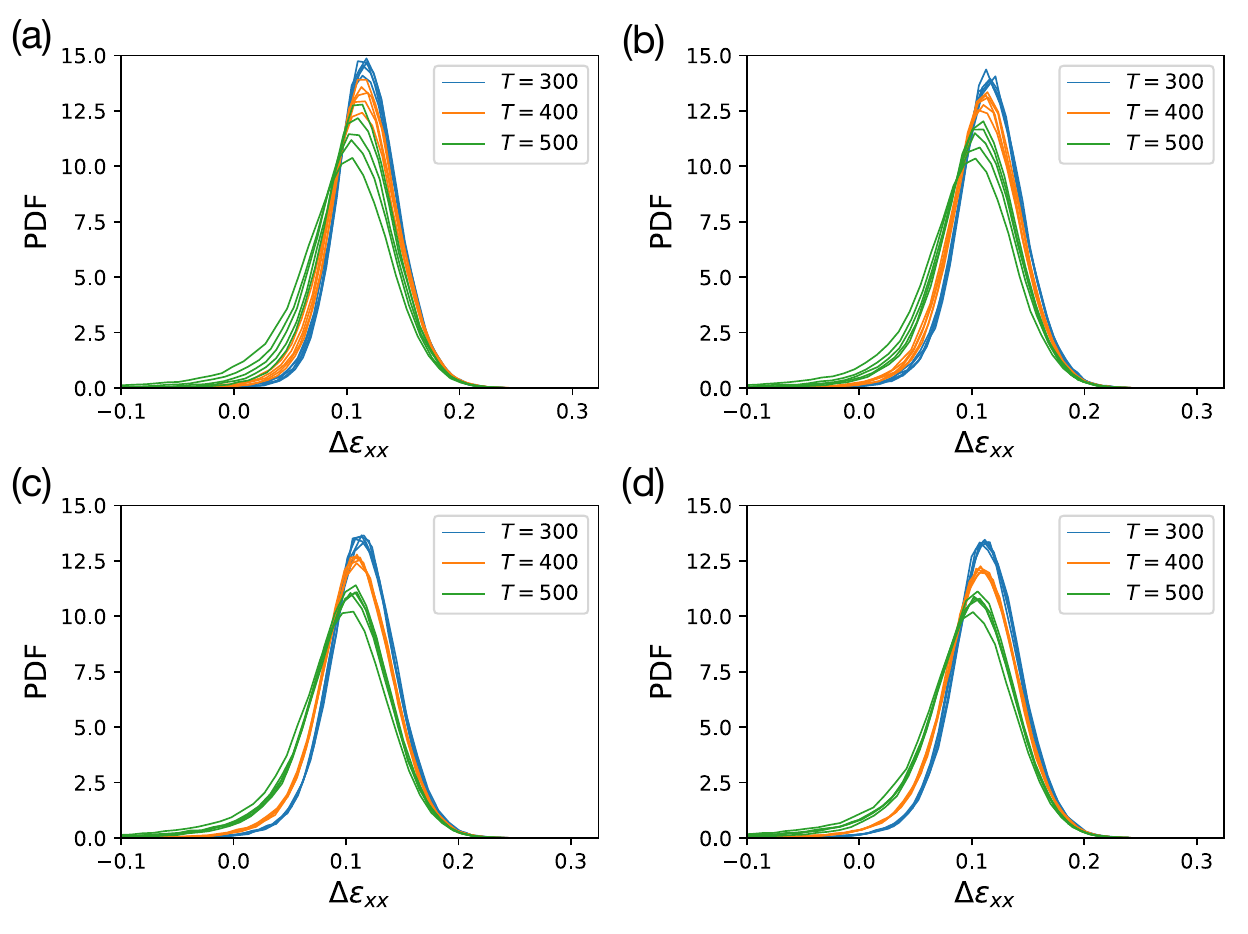}
\caption{The statistical distribution of the eigenstrain $\Delta \epsilon_{xx}$ observed in deformation events in the kMC simulations.
(a) Time-independent STZ model, (b) $\beta = 0.6$, (c) $\beta = 1.0$,
(d) $\beta = 1.3$.}
\label{fig:dist-plast-strain}
\end{figure}

In the present kMC simulations, the eigenstrain produced by activation of an STZ, $\Delta \epsilon^{(m)}_{ij}(\vc{x})$, was randomly assigned from a normal distribution with zero mean and a standard deviation of $0.05$.
However, the strain associated with an event that actually occurs in the kMC simulations is expected to be positive along the external loading direction because the transformation of STZs must relieve the applied stress.
Since the mean of this actual strain $\Delta \bar{\epsilon}$ is one of the important parameters for the theoretical analysis, we investigate the statistical distribution of the strain realized in the kMC simulations.

Because the loading direction of the simulations is the $x$ direction, we show the statistical distribution of $\Delta \epsilon_{xx}^{(m)}$ for all modes in Fig.~\ref{fig:dist-plast-strain}, where we omit the mode index $m$ in the figure.
The figure clearly shows that all the distributions have a peak around $\Delta \epsilon_{xx} = 0.1$.
Based on this result, we adopt the peak value as the mean value of transformation strain in our theoretical analysis: $\Delta \bar{\epsilon} = 0.1$.


\end{document}